\documentclass[preprint,12pt]{aastex}

\shorttitle{Halo stream}
\shortauthors{Yanny, Newberg et al.}

\begin{document}

\title{A Low Latitude Halo Stream around the Milky Way}

\author{
Brian Yanny\altaffilmark{1, \ref{FNAL}},
Heidi Jo Newberg\altaffilmark{1, \ref{RPI}},
Eva K. Grebel\altaffilmark{\ref{MPH}},
Steve Kent\altaffilmark{\ref{FNAL}}
Michael Odenkirchen\altaffilmark{\ref{MPH}},
Connie M. Rockosi\altaffilmark{\ref{UW}},
David Schlegel\altaffilmark{\ref{PU}},
Mark Subbarao\altaffilmark{\ref{UC}},
Jon Brinkmann\altaffilmark{\ref{APO}},
Masataka Fukugita\altaffilmark{\ref{JPG}},
\v{Z}eljko Ivezic\altaffilmark{\ref{PU}},
Don Q. Lamb\altaffilmark{\ref{UC}},
Donald P. Schneider\altaffilmark{\ref{PSU}},
Donald G. York\altaffilmark{\ref{UC}}
}

\altaffiltext{1}{Equal first authors}

\altaffiltext{2}{Fermi National Accelerator Laboratory, P.O. Box 500, Batavia,
IL 60510; yanny@fnal.gov; skent@fnal.gov\label{FNAL}}

\altaffiltext{3}{Dept. of Physics, Applied Physics and Astronomy, Rensselaer
Polytechnic Institute Troy, NY 12180; heidi@rpi.edu\label{RPI}}

\altaffiltext{4}{Max Planck Institute for Astronomy, KonigStuhl 17, D-69117 Heidelberg, Germany \label{MPH}}

\altaffiltext{5}{University of Washington, Seattle, Washington\label{UW}}

\altaffiltext{6}{Princeton University, Princeton NJ 08544\label{PU}}

\altaffiltext{7}{Dept. of Astronomy and Astrophysics, University of Chicago, 5640 S. Ellis Ave., Chicago, IL 60637\label{UC}}

\altaffiltext{8}{Apache Point Observatory, P. O. Box 59, Sunspot, NM 88349-0059\label{APO}}

\altaffiltext{9}{University of Tokyo, Institute for Cosmic Ray Research, Tokyo, Japan\label{JPG}}

\altaffiltext{10}{Department of Astronomy and Astrophysics, The Pennsylvania State University, University Park, PA 16802\label{PSU}}

\begin{abstract}

We present evidence for a ring of stars in the plane of the Milky Way, extending at least
from $l = 180^\circ$ to $l = 227^\circ$ with turnoff magnitude $g \sim 19.5$; 
the ring could encircle the Galaxy.  We infer that the low 
Galactic latitude structure is at a fairly constant distance 
of $R = 18 \pm 2$ kpc from the Galactic Center above the Galactic 
plane, and has $R = 20 \pm 2$ kpc in the region sampled below the 
Galactic plane.
The evidence includes five hundred SDSS spectroscopic radial velocities of stars within
$30^\circ$ of the plane.  
The velocity dispersion of the stars associated with this structure is found to be
$27 \> \rm km \>s^{-1}$  at ($l,b$) = (198$^\circ$, -27$^\circ$), 22 $\rm km\>s^{-1}$ at ($l,b$) = (225$^\circ$, 28$^\circ$), 30 $\rm km\>s^{-1}$ at
($l,b$) = (188$^\circ$, 24$^\circ$), and 30 $\rm km\>s^{-1}$ at ($l,b$) = (182$^\circ$, 27$^\circ$).  The structure rotates in the same prograde direction 
as the Galactic disk stars, but with a circular velocity 
of $110 \pm 25\>\rm km\>s^{-1}$.  
The narrow measured velocity dispersion is inconsistent with power law spheroid or
thick disk populations.  
We compare the velocity dispersion in this structure with the velocity dispersion of stars
in the Sagittarius dwarf galaxy tidal stream, for which we measure a velocity dispersion of 20 $\rm km\>s^{-1}$
at $(l, b) = (165^\circ, -55^\circ)$.  We estimate a preliminary metallicity from the Ca II (K) line and color of the turnoff stars of $[Fe/H] = -1.6$ with a dispersion of 0.3 dex and note that the turnoff color is consistent with that of the 
spheroid population.
We interpret our measurements as evidence for a tidally disrupted
satellite of $2 \times 10^7$ to $5 \times 10^8 \rm M_\odot$
which rings the Galaxy.

\end{abstract}


\section{Introduction\label{intro}}

The Milky Way Galaxy's system of stars is generally divided into
several global components based on broad positional, kinematic and 
chemical abundance information.  Major components are:
thin disk, thick disk, spheroidal halo, and bulge.
There is recent evidence, however, that these basic components 
do not explain all of the stars seen in photometric and spectroscopic surveys.

\citet{lh96} and \citet{phl2001} report a significant asymmetry in the 
number of stars in the thick disk/inner halo from one side of the 
Galactic Center to the other, at angles of $20^\circ - 75^\circ$ from 
a Galactic longitude of zero, at $20^\circ < b < 50^\circ$.  The stars are
approximately 3 kpc from the Sun.  They suggest
this asymmetry could indicate triaxial distributions of thick disk or inner halo 
stars, or could result from interactions of the disk with the Sagittarius 
dwarf or the Milky Way's bar.

Figure 3 of \citet{a01} shows a 35\% asymmetry between the number of stars 
6 degrees above and below the plane, at $l = 240^\circ$.  They do not quote
a typical distance to their stars, but since this study uses brighter
2MASS stars, they must be within a few kiloparsecs of the Sun.

There are several groups who see evidence of a disk-like distribution of 
metal-weak ([Fe/H] $\sim -1.6$) stars, which extends to larger scale heights
and lower metallicities than the canonical thick disk
of \citet{gw85}.  \citet{nbp85} found a set of stars with orbits like disk
stars and metallicities like halo stars, and interpreted these as a
metal-weak tail to the thick disk population.  \citet{mff90} confirmed
and extended this result, and \citet{m93} measured a vertical scale height
of the metal-weak thick disk of 2 kpc (twice the canonical scale height
of the stars of typical metallicity in the thick disk component).

It is not known whether the Galactic disk is formed from separate disk
components, or whether there is a single disk component with a metallicity
and age gradient.  In particular, there is no consensus on whether the 
metal-weak thick disk is part of the proposed Galactic thick disk, or
whether it is itself a separate component \citep{bdrcrfnh2002}.

\citet{gwn2002} find a group of stars, within 3-5 kpc of the Sun at
$l \sim 270^\circ$, which is rotating more slowly about the Galactic center
than would be expected of a thick disk population.  The stars are found
$30^\circ$ above and $45^\circ$ below the Galactic plane (5 kpc apart).
They interpret these stars as originating in a satellite galaxy which merged
with the Milky Way and perhaps could have created the thick disk.

There are other indications that smooth galaxy components
of star distributions in the Milky Way do not yield the complete picture.
Moving groups and halo substructure have been observed from stellar
kinematics \citep{hwzz99,mmh96}.
Tidal disruption of globular clusters
and the Sagittarius dwarf galaxy in the halo of our Milky Way galaxy have
been documented \citep{oetal01,mom98,mskrjtlp99,magc01,v01,retal02,ketal02,bfi02}
in spatial star-density and kinematic studies.

Deep, accurate, multi-color Sloan Digital Sky Survey (SDSS, see York et al. 2000) photometry of blue stars was 
used by \citet{ynetal00}  and \citet{netal02}, hereafter 
designated 
Paper I and Paper II, to probe the structure of the Milky Way's halo over a 
distance range $2 < d < 150$ kpc.  Paper I identified large 
structures (length scales $ > 10\>\rm kpc$) of stars which turned out to be two pieces of the
tidal stream of the Sagittarius dwarf galaxy \citep{iils01}.
Paper II showed the Sagittarius dwarf stream turnoff stars and color-magnitude
diagrams (CMDs), and also identified additional substructure in the Galactic halo.
The most striking discovery was the identification of a large number of blue
stars at low Galactic latitudes, roughly in the direction of the Galactic 
anti-center.

Each overdensity of stars identified in Paper II was designated with a
label of the form S$l\pm b-g$, where ($l\pm b$) are the approximate
center in Galactic coordinates and $g$ is the approximate $g$
magnitude of the F dwarf stars in that overdensity.  Since many of the
structures we are looking at cover large areas of the sky (for example
the stream of the Sagittarius dwarf galaxy), it is likely as we obtain 
additional SDSS imaging data that the many identified overdensities 
will eventually connect to trace large structures.
We concluded in Paper II that the overdensities S223+20-19.4,
S218+22-19.5, S183+22-19.4, and (with less significance) S200-24-19.8
were parts of the same large structure in the sky, since they had
similar CMDs, turnoff colors, and inferred distances.  The sky
positions of these overdensities are spread over $40^\circ$ of
Galactic latitude and $40^\circ$ of Galactic longitude.

In Paper II we explored two possible explanations for the unexpectedly
high concentrations of stars near the Galactic anti-center:  either
the stars are part of the tidal stream from a tidally disrupted dwarf
galaxy, or they are part of the ``metal-weak thick disk" described
above.

One of these structures, at $(l,b) = (198^\circ, -27^\circ)$, was selected for
spectroscopic followup, and we present here radial velocities of
several hundred faint $g \sim 20$ blue stars in that structure.  In
addition, we present serendipitous SDSS spectra of $18.5 < g < 19.5$
stars in several other directions of interest and make a case that
stars in three of these directions are pieces of the same low-latitude stream, at a
distance of 18 kpc from the Galactic center.
We also consider the possibility that the stars are part of a
shell such as that seen around some elliptical galaxies which have
undergone mergers \citep{mc83,mk98,t99}.




The case for the existence of a stream ringing the Galaxy is made 
in this paper as follows:  The instrumentation and observing system 
of the SDSS is reviewed in \S2, and the accuracy with which radial 
velocities obtained with the SDSS are known relatively and absolutely 
is justified in \S4.  In \S3 and \S6, radial velocities of a set of faint blue 
turnoff stars, defined to be stars of spectral type F with absolute 
magnitudes of about $M_g = 4.2 \pm 0.3$, are observed at several positions 
around the sky.
At low Galactic latitudes, these stars at implied distances of 17-20 kpc from 
the Galactic center are shown to have a remarkably narrow
velocity dispersion, inconsistent with a stellar spheroidal halo population.
Combined with the narrow appearance of the stars' faint turnoff in the 
photometric color-magnitude diagram, we argue that the distribution is 
annular in nature.  A comparison to a known stream is discussed in \S5.  
Using the observed velocities and positions of stars over an arc of 
$50^\circ$ on the sky we constrain the possible orbits of stars in this
coherent structure in \S7 and discuss further details in \S8.  Finally, we summarize in \S9.

After submission of this paper, \cite{ietal2003} circulated a paper
showing that the stream extends to the vicinity of the Andromeda galaxy, 
extending the known extent of the structure from $50^\circ$ to $100^\circ$ 
of arc on the sky, as seen from the Sun.

\section {SDSS Technical Details}

A main goal of the SDSS is to obtain the redshifts of nearly one million galaxies and
one hundred thousand quasars. These are obtained with 45 minute long exposures with 
a 640 fiber double spectrograph on the SDSS 2.5m telescope \citep{yetal00}.  The 
targets are selected from SDSS imaging data, and holes at 
the positions of bright galaxies, stars and other targets of 
interest are drilled in 1 m diameter aluminum plates, which cover a $3^\circ$ 
diameter field of view.

During normal SDSS operations, stellar objects are targeted in about
35 to 80 of the 640 fibers on any given plate.  Sixteen of these
targets on each plate are moderately bright objects, called SDSS reddening and
spectrophotometric stars, and are used to calibrate all the objects
on a given plate.  They are generally F subdwarfs with $16 < g <
18.5$.  There are also 10-30 fibers placed on objects of interesting
color, generally blue or red objects off the usual stellar locus,  including blue 
horizontal branch stars, brown
subdwarfs, and white dwarfs.  In addition, about 35\% of fibers placed
on quasar candidates turn out to be stars in the Galaxy rather than
actual quasars.  In this way, a large sample of stellar spectra are
obtained with signal-to-noise greater than 10 to limits of $g \sim
20.0$ over the course of the survey.

Here and throughout the paper, SDSS magnitudes and filters are designated 
without superscripts: $u, g, r, i, z$.  This indicates that the
magnitudes are on the 2.5m imaging camera system, which is
closely tied to the SDSS photometric system. See 
\citet{setal02,figdss96,hetal01} 
for details on the SDSS photometric system and the SDSS photometric monitor.
See \citet{getal98,l03} for details on the SDSS imaging camera system and
imaging processing software.
A subscript `0' on a magnitude or color indicates that it has been corrected 
by applying the full dereddening appropriate for its 
position on the sky.   We correct these magnitudes for reddening 
using $E(B-V)$ from \citet{sfd98}, which has spatial resolution 
of 0.1 degrees, and the standard extinction curve, which for SDSS filters 
yields: $A_{u^*} = 5.2 E(B-V); A_{g^*} = 3.8 E(B-V); A_{r^*} = 2.8 E(B-V)$.  
The astrometry of SDSS objects is good to better 
than 100 mas and techniques used in obtaining this accuracy are 
described in \citet{pmk03}.

During standard SDSS spectroscopic observing, each plate is exposed for 
$3\times 15 $ minutes $= 45 $ minutes on 
the 2.5m SDSS telescope \citep{yetal00} at Apache Point Observatory.  Arc lamps of HgHeNeAr are
obtained before and after each exposure.  Quartz flat exposures 
of the fibers are also obtained.

The data are processed through the standard SDSS spectroscopic 
pipelines (software versions v4\_9\_8 for 2D and v5\_7\_3 for 1D 
processing were used), and result
in a set of 1D spectra as well as a radial velocity for each object 
based on cross correlations against a high signal-to-noise  (S/N) 
template F star.  Additionally, the observed wavelengths and 
equivalent widths of common lines 
like Ca II K and the Balmer lines are measured.
The line positions give a second
measure of radial velocity in addition to the template match, and the
agreement between the two does not conflict with our estimated radial velocity errors
for fainter stars of $\sigma = 16 - 20 \rm \> km\> s^{-1}$, which are derived
in \S4.

The exact selection function of SDSS regular program stellar spectra 
is complex, but the corresponding imaging data are complete 
several magnitudes beyond the spectroscopic sample magnitude limit. Thus,
color and number count information from the imaging can be used to estimate 
completeness of any spectroscopic sample.  The algorithm by which 
a mosaic of $3^\circ$ diameter spectroscopic tiles are selected on the sky
is described in \citet{betal02}.

\section {F star plate 797}

We took advantage of an opportunity when none of the normal SDSS program areas were
visible to use the fibers in one specially designed spectroscopic plate to study the 
dynamics of the stars identified in Paper II as belonging to the overdensity S200-24-19.8.
The name indicates the stars are near $(l, b) = (200^\circ,-24^\circ)$ and the turnoff is at approximately
$g_0 = 19.8$.  With an estimated absolute magnitude of $M_g = 4.2$ for these blue stars,
this puts the overdensity at a distance of 13.2 kpc from the Sun.  


Figure \ref{selectedstars} shows a portion of Figure 15 from Paper
II, which contains SDSS photometry in a $(g-r)_0, g_0$ color magnitude
diagram of stars near $(l,b) = (198^\circ, -27^\circ)$.  Note the faint
turnoff at $(g-r)_0 = 0.2, \> g_0=19.5$.
This turnoff has many more stars than the thin or thick disk
populations at the same implied distance from the Sun along
this Galactic sight-line.

Unlike the usual SDSS science plates, which target about 500 galaxies,
100 QSO candidates and 30 stars per plate, this special plate
797 was arranged to target about 600 blue stellar objects with $19.1 <
g_0 < 20.3$ and $0.16 < (g-r)_0 < 0.3$.  A handful of objects outside of
this range, including eight spectrophotometric and eight reddening stars at
$g_0 \sim 17$ and $g_0 \sim 18.25$ respectively, and a few objects 
with $(g-r)_0 > 0.4$, were also targeted for quality control.
The area in color-magnitude space where the spectroscopic targets on plate 797 were 
selected is indicated by the triangular points in Figure \ref{selectedstars}.

Plate 797 was exposed for a total of 7200 s on the nights of 2001
December 19 and 20 with the SDSS spectrographs, and processed with standard
SDSS software for spectroscopic reductions.  This produced
a set of fully extracted, wavelength and flux
calibrated 1D spectra.

Each of the $\sim$ 600 F-star spectra was examined individually.  Since no
$(u-g)_0$ color cut was applied to the selection of objects on this particular
plate, spectra of numerous hot objects, unrelated to the target F turnoff 
stars, were observed.  Spectra of about 50 quasars and numerous 
white dwarfs, as well as spectra with S/N too low to measure an 
accurate radial velocity, were rejected.  Rejecting objects with 
S/N less than about 5:1
 effectively sets our faint limit for obtaining a good radial velocity 
on this plate at about $g=20.5$.
This left 392 good F star spectra, including the brighter
reference F and G stars at $17.0 < g_0 < 18.5$.  

A sampling of the
good plate spectra shows that the typical S/N for these 
spectra ranged from about 10 at $g_0=19.5$ to about 5 at $g_0=20.1$.
Individual positions, magnitudes, colors, SDSS IDs,
radial velocities, and the equivalent width of the 
Ca II K $\lambda 3933$ \AA\ line ($W_K$) for each good spectrum 
are listed in Table 1 (full table is ascii text).  Table 1 lists for 
each object:
right ascension, declination (J2000), SDSS imaging identifying code 
(format run-rerun-camcol-field-id), spectroscopic identifying 
(format plate-mjd-fiberid), $g_0$ magnitude, $(g-r)_0$ color 
and $(u-g)_0$ color,
the heliocentric radial velocity in $\rm \>km\>s^{-1}$,
$W_K$, a color cut
flag, where a 1 indicates the object is one of 327 in the color magnitude box:
($0.158 < (g-r)_0 < 0.3 ,\> 19.1 < g_0 < 20.3$), and the $E(B-V)$ from \citet{sfd98} used to deredden each object.
Figure \ref{rvhist} shows the histogram of observed radial velocities
for these 327 stars in the color-magnitude box of plate 797.

One can compare the radial velocity histogram of the stars in Figure
\ref{rvhist} with our expectations for the radial velocity
distribution of standard Galactic components.  Each component (power
law spheroid, thick disk, and thin disk) is assigned a nominal
Galactic rotation velocity in the solar neighborhood and a velocity
dispersion ellipsoid.  For each component, we calculate the radial
velocity, velocity dispersion, and density of stars at the position
and distance of the stars in Figure \ref{rvhist}.  For the thin disk, 
we adopt
\[<v_{rot},\sigma_{U}:\sigma_{V}:\sigma_{W}> = <220, \>38:25:20>,\]
in units of $\rm km\>s^{-1}$;  for the thick disk, 
\[<v_{rot}, \sigma_{U} : \sigma_{V} : \sigma_{W}>  =  <170,  \> 60:45:40>;\]
and for the halo,
\[<v_{rot},\sigma_{U}:\sigma_{V}:\sigma_{W}> = <0, \> 130:100:85>,\]
in the same units.  These parameters are taken primarily from 
the compilation in Allen (2000), p. 479.  However, we adopted a 
non-rotating halo component and set the rotation velocity of the 
thin disk to the rotation velocity of the stars in 
the solar neighborhood.  The results of fitting this
model to the data are not sensitive to small differences in our adopted
parameters.  


The Galactic models are calculated as follows:
Since the Galaxy has an essentially flat rotation curve exterior to
the solar position out to some large distance, the rotation speed of each 
component in the solar neighborhood is used as the estimate of 
the rotation curve at the
position of the stars in plate 797 ($d_\odot = 13$ kpc, $r_{GC} =
20$ kpc).  First, we convert position in the Galaxy $(l, b, r)$ to
standard Galactic Cartesian coordinates $(X, Y, Z)$.  From the X and Y
positions, we calculate the components of the rotation velocity in the
$(\hat{x}, \hat{y})$ directions:
\[\theta = \tan^{-1}{Y/X},\]
\[v_x = v_{rot} \sin{\theta},\]
\[v_y = -v_{rot} \cos{\theta}.\]
The position and velocity of the Sun are taken as: $(X, Y, Z) = (-8,
0, 0)$, $(v_x, v_y, v_z) = (10, 225, 7)$ where distances are in kpc
and velocities are in $\rm km\>s^{-1}$.  If $\hat{r}$ is the unit vector in the
direction from the Sun to the portion of the Galaxy which we are
studying, then the radial velocity we expect to observe from this
component is: 
\[v_R = ( \vec{V}-\vec{V_\odot} ) \cdot \hat{r}.\]
From the previous equation, we deduce that the velocity
dispersion in the direction ($\hat{r}$) is given by:

 \[ \sigma_{rv} = \sqrt{(\hat{x}\cdot\hat{r})^2{\sigma_x}^2 + (\hat{y}\cdot\hat{r})^2{\sigma_y}^2 + (\hat{z}\cdot\hat{r})^2{\sigma_z}^2 }.\]

It is in principle very difficult to calculate either the fraction of the
stars which should belong to each component of the Galaxy or to
additionally calculate an overall normalization.  This is because one
requires not only all of the parameters in the density model for each
component, but also the component normalization (usually in the solar
neighborhood) of the star counts, and these must be {\it for the
colors of the stars used to select the spectroscopic sample.}  If
there are color gradients in any component, then these must be modeled
as well.  Since
we use a blue color cut here, in fact bluer than most of the stars in
the thick disk component, we assume the normalization between the
thick disk and halo might not be the same as if we had used a redder
cutoff.  In spite of this known difficulty, we estimate the number 
of stars expected from each component using the following method.

We adopt a model for the density distribution of each Galactic component
which is similar to those used in Figure 20 of Paper II.  That Figure plots
number counts of observed stars around the celestial equator against
models of the number counts expected from standard thin disk, thick disk 
and halo models.  The exponential thin disk has scale height 0.25 kpc 
and scale length 2.5 kpc.  For the (exponential) thick disk, we use a scale
height of 1.0 kpc, a scale length of 3 kpc, and a normalization of one
in 30 of the thin disk star counts in the solar neighborhood (but see Chen et al. 2001).  

We use a spherical, power law halo distribution with $\alpha = -3.5$ and a
normalization of one in 500 of the thin disk star counts in the solar
neighborhood \citep{bs84}.  We fixed the overall normalization 
of the model to match Fig. 20 of Paper II;  the star counts in the magnitude range
$16.5 < g_0 < 17.5$ (with assumed absolute magnitude $M_{g} = 5.0$, since
these stars were selected from a redder population than the stream selection)
in the direction ($\alpha_{2000},\delta_{2000}$) = (240$^\circ$, 0$^\circ$) 
were fixed to match our measured 388
per square degree.  Note that the stars in Fig. 20 of Paper II are
selected with $0.2 < (g-r)_0 < 0.5$; normalizations used in this
model will produce the number of stars expected per square degree, in
the color range $0.2 < (g-r)_0 < 0.5$, in
a given distance range (calculated from an apparent magnitude range
and estimated absolute magnitude for the typical star in the sample).
For reference, $(g-r) = 0.98(B-V) - 0.19$ \citep{setal02}.
The normalization of the the Galactic components is determined by assuming
an absolute magnitude of $M_g = 4.3$ (rather than 4.2 for the plate 797 stars), 
since the average spectrum is from a star slightly redder than 
the turnoff stars.  The model integrates the star
counts expected over the magnitude range $18.9 < g < 20$, which is the
selection box for the spectra.

These models predict that there are 5 thick disk stars 
and 131 halo stars per square degree with $0.2 < (g-r)_0 < 0.5$ in the direction
and magnitude ranges of the stars in plate 797.  Since these models
predict that there should be a factor of $10^5$ fewer thin disk stars
than thick disk stars, we will assume that there are no thin disk stars
in our faint blue sample.

We now estimate the number of blue stars ($0.18 < g-r < 0.28$) we
expect to find in the same volume.  Figure 15 of Paper II shows a
Galactic color magnitude diagram in a direction of sky including
that towards plate 797; it is a full version of the color magnitude
diagram shown in Figure 1.  From Figure 15 of Paper II, we
estimate that the slope of the thick disk main sequence is 6.6 in the
$g_0$ vs. $(g-r)_0$ diagram.  From that same diagram, we attempt to select a
sample of mainly thick disk stars that are all at the same distance
(and thus in the same volume).  From the SDSS
stars in the area of the sky described by $-1.25^\circ < \delta_{2000} < 1.25^\circ$,
$70.5^\circ < \alpha_{2000} < 73.5^\circ$, we select the stars with $g_0 > 6.6 (g-r)_0 +
15.06$ and $g_0 < 6.6 (g-r)_0 + 15.36$.  These latter cuts have the
effect of selecting stars with $17.7 < g_0 < 18$ at $(g-r)_0 = 0.4$.  Of
these stars, 97 are in the color range $0.18 < (g-r)_0 < 0.28$ and 664
are in the color range $0.2 < (g-r)_0 < 0.5$.  From this, we estimate
that there are (very roughly) 15\% as many blue stars in the thick disk 
as there are redder stars at the same distance.

We use the same logic to select a set of primarily halo stars, as
judged from Figure 8 of Paper II.  We choose stars from the area of
sky: $-1.25^\circ < \delta_{2000} < 1.25^\circ$ and $240^\circ < \alpha_{2000} < 243^\circ$.  Using the same
slope for the main sequence, but shifting the magnitude range so that
$20.5 < g_0 < 20.8$ at $(g-r)_0 = 0.4$, we find 270 stars in the color
range $0.18 < (g-r)_0 < 0.28$ and 1420 stars in the color range $0.2 <
(g-r)_0 < 0.5$.  From this we estimate that there are roughly 19\% as 
many blue stars in the halo population as there are redder stars at the same
distance.

The predicted contributions to the radial velocity distribution from the thin disk,
thick disk, and power law spheroid components are shown in Figure \ref{rvhist} in red, green,
and blue, respectively.
The mean and standard deviation of the model Gaussian distribution are derived
from the models of each stellar component, and the normalizations are estimates
derived from the models and our measurements of stellar distributions in the Galaxy,
as described above.
The sum of all distributions is shown by a
heavy, black, dashed line.  These known components only account for 157 of the
325 stars shown in the Figure.

We now estimate the mean and dispersion of the ``extra" stars.  Since the only
components which contribute a significant number of stars to the diagram are
the power-law spheroid and the additional population whose properties we wish
to measure, we used a maximum likelihood fit to the data in which the model included
only two Gaussian distributions.  The
mean, sigma, and fraction of the observed stars in the extra component
were allowed to vary so that the reduced chi-square fit between the
model and the data was minimized (reduced chi-squared = 1.33).  We fixed the 
total number of stars as the observed number
of stars, but allowed the relative amplitudes of the two Gaussian components
to vary, since the total number of stars in the spheroid population is rather
poorly known.  The minimum chi-squared was achieved if the extra component was
assigned a mean of 74 $\rm km\>s^{-1}$ and a sigma of 34 $\rm km\>s^{-1}$.  The 
number of stars assigned to the spheroid population, 153, was very close 
to the 174 that were calculated from our standard galaxy model, giving us
added confidence in that estimate.  The thinner black curve in 
Figure \ref{rvhist} shows the extra distribution of stars, and the thicker 
black curve shows the sum of the extra stars with the standard model.  The 
thick black curve shows reasonable agreement with the data.  

One could imagine adjusting each of the model parameters (i.e. increase the normalization
of the halo component), or adding additional
model components (i.e. a second inner halo or metal weak thick disk component) in an 
attempt to fit the data without the need for this ``extra"
component.  However, one must keep in mind
the following points when imagining an expanded model:

(1) We do not expect {\it any} thin disk stars in our spectroscopic sample, even
if we tweak the scale lengths and normalization to other components within
reasonable bounds.  Warps or flares of a thin (or thick disk) do not work,
as they do not provide enough stars this far from the Galactic center.
A spiral arm at this distance from the Sun would be 4 kpc above the
Galactic plane.

(2) The velocity dispersion of the star sample is {\it much} narrower than that of the
power law spheroid, and is significantly smaller than the local thick disk
velocity dispersion.

If one wanted to fit the radial velocity distribution with a smooth distribution of
stars, then a distribution with much larger scale lengths than the thick disk
would be required (scale height 2 kpc, scale length 8 kpc, as described in 
Paper II).  However, larger scale lengths generally lead to larger dispersions,
which makes the agreement with the data in this respect problematic.

We will show in the next section that a typical instrumental error in the radial velocity dispersion
for this data is $\sigma = 20 \rm \> km\> \>s^{-1}$.  Using this, we estimate that the intrinsic velocity
dispersion of this extra component is 27 $\rm km \>s^{-1}$.  The mean radial velocity of this
component is 74 $\pm 5 \rm \>km\>s^{-1}$.  The error in mean is calculated from the uncorrected
dispersion divided by the square root of the number of measured stars in that component and including the 4 $\rm \>km\>s^{-1}$ plate-to-plate absolute
error (see \S4 below) in quadrature.

Because the stars are at
$l=198^\circ$, relatively near the anti-center, one has only a limited kinematic lever arm for
interpreting radial velocities as projected circular velocities as seen from the Sun.   The
models in Figure \ref{rvhist} calculate these projections, and find that thin disk
($ v_c= 220 \rm \>km \>s^{-1}$) stars would have apparent
radial velocities of $\rm +49 \>km \>s^{-1}$, thick disk stars (rotating in a
cylinder with $ v_c = 170 \rm \>km \>s^{-1}$) would show motions 
of  $+54 \rm \>km \>s^{-1}$ and a stationary halo would show radial velocities 
of $+73 \rm \>km \>s^{-1}$.  Any distribution of stars with significant observed mean
radial velocity greater than $73 \rm \>km \>s^{-1}$ which is also in a co-planar 
circular orbit would then be moving retrograde with respect to the disk.

These radial velocity estimates depend somewhat on our estimate of the 
distance to the star stream.
The distance to the stream is determined from the distance modulus to 
the turnoff stars, which depends on the presumed absolute magnitude of 
these stars.  The error in the distance modulus, composed of approximately 
equal parts error in the determination of the magnitude of the turnoff 
and in the error of the determination of the absolute magnitude of 
a turnoff star, is about 0.3 magnitudes, or 15\% in distance
from the sun (Paper II).  The absolute magnitude of the turnoff
stars was determined from the magnitude difference between the horizontal
branch and main sequence stars in the same color range in the Sagittarius
dwarf tidal stream (measured in SDSS filters).  The result was confirmed
by computing the absolute magnitudes of stars of similar color
in the globular cluster Pal 5, as measured in the SDSS (Paper I).
The metallicity of Pal 5 is $[Fe/H] = -1.5$, whereas the metallicity of
the Sagittarius dwarf tidal stream is unknown due to age uncertainty.
The Pal 5 metallicity is similar to that derived for stars on 
plate 797 (see \S8 below) and thus the adopted distance moduli are 
appropriate within the quoted systematic error.  The one sigma error 
in distance modulus gives us a distance 
range of 11.3 -- 15.0 kpc (from the sun).
There is, in addition, uncertainty in the intrinsic radial thickness of 
the star structure, which has approximately the same amplitude, ($< 4$ kpc), 
and is discussed below.

Since we are looking close to the Galactic anticenter, the assumed distance 
to the stars is of less importance.  The measured circulation velocity of 
the stars, assuming they are in a circular orbit, is 
$-5 \pm 44 \pm 1 \pm 30 \rm \>\>km\>\>s^{-1}$, where the first error
term is the one sigma contribution from the error in radial velocity 
determination, the second contribution to the error is from the error 
in distance, and the third contribution to the error is from the difference 
in radial velocity across the plate.  For a circularly rotating
group of stars, the radial velocities should shift by about 11 
$\rm \>km\>s^{-1}$ from one side of the plate to the other side.  This 
shift is discernable in the radial velocity data.  This effect
increases our measured radial velocity dispersion by a few 
$\rm \>km\>s^{-1}$.  Even though
formally we derive $v_c = -5 \pm 54 \>\rm \>km\>s^{-1}$ (no net rotation) 
for stars along this line of sight, we will demonstrate in \S6 below that the 
circulation velocity of stars above the Galactic plane in this structure 
are consistent with a prograde rotation.

\section {Standard Plate 321}

Confirmation of the radial velocity zero-point and rms error in the
SDSS radial velocities was determined by examining a special SDSS
calibration plate.  This plate, designated plate 321, specifically
targeted the open cluster M67, where radial velocities are known to
accuracies of $\rm 1\> \>km\>s^{-1}$ \citep{mlgg86}.  Exposures of
60s, 120s and 240s were obtained of this field on 2000 March 9, and
processed in a manner identical to that of plate 797.  The resulting
80 spectra typically have S/N of 50:1.
With this large signal-to-noise ratio, the error in the radial velocities
is determined from the resolution of the spectrographs and systematics
of the wavelength calibration.

Figure \ref{stdplot} shows the difference between the
SDSS radial velocities and the 
\citet{mlgg86} catalog velocities of bright blue stars in the open
cluster M67 vs. the fiber number.  Each plate has 640
fibers, divided into two `left/right' halves of 320 fibers each.  The SDSS
instrumentation consists of a double spectrograph, with 
two blue and two red cameras. The fibers from the left half (numbered 1:320) of each
plate are fed to one spectrograph; those from the right half (number 321:640) to another. 
From Figure \ref{stdplot} we see that there is a systematic difference of
$4 \rm \>km\>s^{-1}$ between
the wavelength calibration on the left and right spectrographs.
This difference results from separate wavelength calibrations for each spectrograph,
based on arc lamp exposures before and after the target exposures. 
It is expected that this systematic would also be present from 
plate to plate, with an amplitude similar to $4 \rm \>km\>s^{-1}$.
This systematic is very small compared with the instrumental resolution 
of the SDSS spectrographs of 150 $\rm \>km\>s^{-1}$.  

If the radial velocities of the left side of the spectrograph are adjusted
by $4 \rm \>km\>s^{-1}$, then the standard deviation of the radial velocity
errors is $11 \rm \>km\>s^{-1}$. 
Because the exposure times on plate 321 are quite short, the night sky lines used 
to assist in setting the wavelength scale in each fiber are underexposed relative 
to normal SDSS plate, and thus the scatter in radial velocities can
in principal be somewhat better than $11\rm \>km\>s^{-1}$ for longer exposures.
The resolution and flexure of the spectrographs limits the accuracy with 
which we can determine radial velocities 
to about $7\rm \>km\>s^{-1}$ for typical SDSS
exposure times.  This number is derived from internal SDSS diagnostic testing.

We further check this radial velocity error empirically by using SDSS
`Quality Assurance' spectra of blue stars.  `Quality Assurance' spectra are repeats
of SDSS targets previously observed on other spectroscopic plates, which are taken
to assess the overall quality of the spectroscopic program.  We selected 146 stellar
objects with $0.15 < (g-r)_0 < 0.36$, and $18.9 < g_0 < 20$, which were observed on two
different plates.  Figure \ref{rverrplt} shows 
the errors in radial velocity determination for these stars, which have a typical
S/N of 10.  The dispersion in the pairs of measurements, 
divided by $\sqrt{2}$, is $16 \rm \>km\>s^{-1}$; we
adopt this value for correcting our observed dispersion to an intrinsic
dispersion for objects with S/N $\sim 10$.

From these measurements, we generate a (somewhat ad hoc) formula to estimate
the accuracy of individual radial velocity determinations as a function of S/N:

\[\sigma = \sqrt{7^2 + \left(\frac{150}{(S/N)}\right)^2} \rm \>km\>s^{-1}.\]
Plate 797 observations have a typical S/N of 8, rather than 10, so we 
adopt an instrumental spread of 20 $\rm km\>s^{-1}$ for objects
on this plate when correcting the observed dispersion in Figure \ref{rvhist} from
34 to 27 $\rm km\> s^{-1}$.

\section {Comparison: Dispersion of stars in the Sagittarius Stream South}

Since the velocity dispersion of the stars towards the anticenter is much narrower
than expected for the smooth Galactic components represented at 5 kpc above the plane, we ask
whether it is similar to the dispersion of a tidal stream.  
At least two works have estimated the dispersion within the Sagittarius dwarf tidal stream:
\citet{mskrjtlp99} find a velocity dispersion of 27 $\rm \>km\>s^{-1}$.
\citet{dpetal01} measure a dispersion of 31 $\rm \>km \>s^{-1}$.  

Fortuitously, the SDSS has also obtained a sample of
stellar spectra in the vicinity of the Sagittarius stream where it
crosses the Celestial Equator.  Sagittarius South stream stars (south
here refers to south of the Galactic plane on the celestial equator,
see Paper II) are seen with $10^\circ < \alpha_{2000} < 45^\circ$, $-1.5^\circ <
\delta_{2000} < 1.5^\circ$ (see Figures 1 and 7 of Paper II).  The turnoff of
the stream is at magnitude $g \sim 21$, which is somewhat fainter than
the limit of ordinary SDSS spectra; however, we have spectra of a number of
blue stars with mags $17.8 < g_0 < 20.2$ which are candidate blue horizontal
branch and blue straggler stars from the Sagittarius stream.
The color magnitude diagram of the stars in the direction of the Sag. south
stream selected area is shown in Figure \ref{cmdsagbox}.  The black box in the figure
was used to select candidate Sagittarius stream stars with $-0.3 < (g-r)_0 < 0.2$.
A histogram of the radial velocities of 306 selected stars
is shown in Figure \ref{rvsagsouth}.

As for the data from plate 797, we show the distribution of stars expected
in Figure \ref{rvsagsouth} from a standard Galactic model, and fit the mean
and standard deviation of the excess stars.  Using the same Galactic model as for the plate
797 data, we find that there are very few thick disk stars expected in the sample (compared
to the expected number of stars from the power law spheroid distribution), and a
vanishingly small number of thin disk stars are expected.  We did not attempt to directly calculate the
normalization of the standard model star counts, since the selection function for these
stars is so poorly known.  However, the number of stars assigned to the smooth components
using maximum likelihood techniques
is similar to the number of spectra in this color and magnitude range taken from
a similar area of sky at about the same Galactic latitude, but in a region that does not
contain the Sagittarius stream.

As before, we fit the sum of two Gaussian distributions to the radial velocity histogram.
The mean and standard deviation of the spheroid population is calculated from the model, and
the sum of the number of stars in both distributions is fixed at the number of stars observed.
The minimum reduced chi-squared of 1.21 was achieved with a mean radial velocity in the
stream of -111 $\rm km\>s^{-1}$, a velocity dispersion in the stream of 21 $\rm km \>s^{-1}$, and 119 of the 306
stars assigned to the stream population.  As these stars are bright, the observational error 
in the RVs is about 7 $\rm km\>s^{-1}$.
Correcting for this error, we quote a true dispersion of the Sag stream 
at this point in space of only 20 $\rm km\>s^{-1}$.  The mean radial velocity is 
an excellent match to the Sagittarius stream models of \citet{ilitq01}.  
The dispersion of the stars observed is similar to, but even
narrower than, previous determinations in 
the literature (which are taken at different points in the stream orbit).

\section{Kinematics of the low latitude stream from other SDSS lines-of-sight}

The spectroscopic observations of faint blue turnoff stars on 
plate 797 showed a narrow dispersion of about 27 $\rm km\>s^{-1}$ in
this anti-center direction below the Galactic plane at
$(l,b) = (198^\circ, -27^\circ)$.  Assuming these stars represent a coherent structure in space and velocity, we now ask if this structure is related to the
overdensities of stars detected photometrically in Paper II
in several places above the Galactic plane.   We address this
by examining the radial velocities of stars from SDSS (regular program)
spectra which are available at low latitude in the same color-magnitude
box as those of plate 797.   The SDSS survey ellipse, in an attempt
to avoid regions of higher than average Galactic reddening,
dips down to lower Galactic latitude towards the anti-center at
the beginning of most survey stripes \citep{setal01}, and this
is where we find most of our spectra.  

%
%
%
%
%
%
%


The main areas of interest where SDSS stellar spectra were available are near
the areas from Paper II designated:  
S223+20-19.4, S183+22-19.4, and S167-54-21.5. 
A typical turnoff star in these structures has $g_0 = 19.4$, implying a 
distance to the Sun of $\sim 11 \> \rm kpc$.  

We note that Figure 16 of Paper II has an error in the labeling
of the upper Figure showing S182+22-19.4.  The label in the lower 
left corner of  Figure
which reads $(\alpha_{2000},\delta_{2000}) = (125^\circ,50^\circ)$ was improperly corrected
for the off-equatorial cos($\delta_{2000}$) and should
read $(\alpha_{2000},\delta_{2000}) = (150^\circ,50^\circ)$.

A color magnitude diagram of SDSS photometry of stellar objects 
in a box with bounds 
($177^\circ < l < 193^\circ, 20^\circ < b < 30^\circ$) 
is shown in Figure \ref{cmd30-34}.
Quasars have been rejected by using the color selection $(u-g)_0 > 0.4$.  
A mixture 
of thin and thick disk turnoff stars are seen in Figure
\ref{cmd30-34} at $(g-r)_0 = 0.4$ down to magnitude about $g_0 \sim
19$.  Bluer halo (spheroid) and other stars are seen to the left of
this.   A `faint blue turnoff' box with $18.7 < g_0
< 20.1$ and $0.15 < (g-r)_0 < 0.36$ has been outlined with a white box
on the Figure.   Available SDSS spectra with photometry in this box 
are selected.

We were able to find three small areas of sky below Galactic latitude $l= 30^\circ$
which contained a significant number of these bluer SDSS stellar spectra.
In the S223+20-19.4 region, SDSS spectra were selected in a box bounded by
$130^\circ < \alpha < 136^\circ , -1.25^\circ < \delta < 3.75^\circ$.  We name
these groups of spectra similarly to the way we named structures in Paper II,
with Galactic longitude, latitude, and turnoff magnitude: Sp225+28-19.6.
Within region S183+22-19.4, we pick two sets of SDSS spectra,
those in the box $179^\circ < l < 185^\circ, 25^\circ < b < 30^\circ$ are designated
Sp182+27-19.4; and those in the box $185^\circ < l < 191^\circ, 21^\circ < b < 26^\circ$ 
are designated Sp188+24-19.3.

We also selected stellar spectra, using the same color and magnitude criteria,
from the high absolute latitude field which contains the Sagittarius stream (south) 
as a ``control."  The selection region is outlined by a white box in 
Figure  \ref{cmdsagbox}.  Stellar spectra from this high (absolute) latitude field are
selected near S167-54-21.5 in the box $25^\circ < \alpha < 50^\circ , -1.25^\circ < \delta < 1.25 ^\circ$. 
Along with the plate 797 stars 
with $70.5^\circ < \alpha < 73.5^\circ, -1.25^\circ < \delta < 1.25^\circ$
our sky coverage includes five areas with SDSS spectra: four at low latitude and
one control field at high latitude.







Samples of spectra from each type of dataset (a plate 321 standard, a Sagittarius
stream blue horizontal branch (BHB) star, a Sp225+28-19.6 Monoceros stream turnoff star, and two stars from
plate 797 at the high and low S/N limits) are presented in Figure \ref{samplespecs}.  
The spectra shown are smoothed over three pixels (210 $\rm km\>s^{-1}$).

Figure \ref{rvhists} shows four radial velocity histograms of faint blue
stars with the same colors and magnitudes as those observed on Plate 797:  The 
upper left hand panel shows stars from Sp225+28-19.6; the upper right and
lower left hand panel shows from the Sp182+27-19.4 and Sp188+24-19.3 areas, 
respectively, and the lower right hand panel shows those from the direction
of the Sagittarius south stream.
All panels except the Sagittarius panel show 
significant numbers of clumped radial velocities with dispersions far smaller 
than one expects for a spheroidal 
population, thus arguing that the population represented here is 
not associated with the stellar spheroid.  
Models of the expected number of stars
due to the standard disks and spheroid from our Milky Way are determined
in a similar fashion to that of Figure \ref{rvhist} and are also shown in each
panel of Figure \ref{rvhists}.

For the low latitude fields, the model predicts ten times as many halo stars as
thick disk stars in each sample.  In the high latitude field, it predicts a hundred
times as many halo stars as thick disk stars.  Therefore, we are again 
justified in using maximum likelihood techniques to fit only two Gaussians - 
one for the halo and one for
the extra distribution.  The technique is identical to the one used in 
fitting the Sagittarius stream.  The mean and dispersion of the 
excess stars is indicated in Figure \ref{rvhists} and in Table 2.


The first five columns of Table 2 list the region name, right ascension and declination (J2000)
of the center of the region, and Galactic $l$ and $b$, for the four regions where spectra were obtained.
Column six lists the average radial velocities with errors.  According to the maximum
likelihood fits, there were 36, 34, and 37 stars in the extra component in the Sp225+28-19.6, Sp182+27-194, and Sp188+24-19.3 directions, respectively.  The errors in the average radial
velocities were determined from the measured dispersion divided by the square root of the
number of stars with 4 $\rm km\>s^{-1}$ plate-to-plate systematic error added in quadrature.  
Since the spectra in Figure \ref{rvhists} have a typical signal-to-noise of 10, we adopt an instrumental
error in radial velocity of 16 $\rm km\>s^{-1}$.  The 
1D velocity dispersions of stars in column 7 are the measured dispersion with 16 $\rm km\>s^{-1}$
subtracted in quadrature.  Columns 8 and 9 list
the $g_0$ magnitude and $(g-r)_0$ color of stars at the turnoff of the faint blue stars.

The last column of Table 2 gives the derived rotation velocity ($v_c$) of the 
group of stars in each set with errors, assuming that they are orbiting the 
center of the Galaxy in a circle.  Although individual stars in the structure must
orbit the center of mass of the Galaxy (not in cylindrical orbits),
one can still use the concept of a coherent circular velocity similar
to that used for the thin and thick disks.  The quoted error on this circular velocity 
is derived by combining the following factors in quadrature: 1) A one sigma error 
on the observed radial velocity; 2) A 15\% error in the distance to the stars, based on 
uncertainty in the measured turnoff magnitude; 3) A $1^\circ - 2^\circ$ error 
in the position of the stars in Galactic longitude (the error due to small 
changes in latitude is much smaller) to allow for the fact that the 
region over which the stars were observed has finite size (and in 
fact is up to a $5^\circ \times 5^\circ$ box).  The values of $v_c$ for 
Galactic longitudes very near the anti-center have, of course, the largest 
errors; the measured circulation velocity for Sp182+27-19.4 is meaningless, since
$<l> = 182^\circ$, and is not calculated.  We note that the quoted $v_c$ values are
consistent within only $2.4\sigma$ of each other, suggesting that the assumption
of a coherent circular velocity model is too simplistic.

In each case, rejecting now the higher absolute latitude 
Sagittarius stream region, we note an excess of stars clumped in radial
velocity with a dispersion consistent with that of the stars on
plate 797.  We suggest that all these stars are related, and may be
moving coherently around the Galaxy at a distance of 18 kpc from
the Galactic center.

\section {A Simple Low-latitude Stream Model} 

In order to explore the possibility that four of the regions surveyed, 
which differ in position
on the sky by many tens of degrees and by implication many kiloparsecs,
are part of one low latitude stream, we have generated a simple model
which fits kinematic and photometric data in these regions 
of interest.  The model simply integrates the equations of motion 
in a static potential.  Two potentials were tried.  The first is a logarithmic 
potential of the form 
$\Phi \sim \ln(s)$ \citep{r80}, where $s^2 = R^2 + Z^2/q_p^2$, 
$R^2 = X^2+Y^2$, $q_p$ is the potential flattening in the Galactic
rectangular coordinate system (X,Y,Z), and the Sun is at $(-8,0,0)$,
rotating in the disk with space velocity $(10, 225, 7) \> \rm
km\>s^{-1}$.  The second model is the disk-halo potential of
\citet{mrs81}, which also has a flattening parameter $q = 1/(1+e)$.
In this second model, $q$ has a somewhat different interpretation 
than $q_p$; it indicates a bulge fraction in a bulge and disk system.

We experiment with different values of $q_p$ and an initial position
and velocity vector for a test particle in the proposed satellite until 
we achieve an orbit which both passes approximately 1 kpc closer to the plane in Z
than the space locations of the four regions of interest and which
also reproduces the average observed radial velocities of the turnoff
stars listed in Table 2 in each of the four regions.
We match our model to Z positions
lower than those of our observed data to reflect the fact that we never see the peak of the star counts, which must be below our observation limits of 
$|b| < 15^\circ$. All these constraints are not 
sufficient to completely constrain the potential,
given the errors, but they do provide a plausible simultaneous fit to 
the photometric and kinematic data in hand.

Figure \ref{pltorbit} shows one stellar orbit which passes
close to the positions of the detections of the stream 
at low latitude.  The model parameters and goodness 
of the fit are listed in Table 3.  In this table $(x_0,y_0,z_0)$ and $(vx_0, vy_0, vz_0)$
are the initial position and velocity of a star in a stream orbit.  The initial position
was chosen to pass through an observed piece of the stream, and the initial velocity,
when integrated, minimizes the least square distances to three other positions
near where data are observed while simultaneously fitting for the
observed radial velocity when the stream particle makes its nearest approach over three
periods.

We note here how we use the agreement in radial velocity between the model
and the observations near $l= 225^\circ$ to constrain the `direction' of the stream.
Towards $l=225^\circ$, the upper left hand panel of Figure \ref{rvhists} shows the stream stars orbiting with 
an average radial velocity ($103 \rm \>km \> s^{-1}$) are close to that of a projected thick disk ($94 \rm \>km \> s^{-1}$), and  significantly
far from that of a stationary halo (143 $\rm km\>s^{-1}$).  Giving this
value the most weight in an averaging of the circular velocities of Table 2,
we thus derive that if the structure seen in Figure \ref{rvhist} and the 
first three panels of Figure \ref{rvhists} are one coherent structure 
moving circularly then it is circling the center of the Milky 
Way with a systemic velocity of $110\pm 25 \rm \>km\>s^{-1}$ in the prograde
direction (with the Galactic disk).  The observations 
near $l=180^\circ$ are less useful because of the diminished lever arm 
so close to the Galactic anti-center, but are not inconsistent with this result.

The model shown here is admittedly very naive, with
no tidal effects or dynamical friction included, but it does
fit the observations, both in velocity and position to within the errors.

Figure \ref{radec} shows where the stars on the stream in the simple model of Table 3 would appear in
the sky.  This diagram is designed to be similar in appearance to
those of \citet{ilitq01} for the Sagittarius stream.  Note how this
stream, which we label the Monoceros stream after the constellation in
which the S225+28-19.4 region lies, remains at $|b| < 30^\circ$ along its
entire orbit.  
Note that elements of the stream in directions towards the Galactic
center may be more difficult to detect than those towards the Galactic
anticenter since they are further away from the position of the Sun,
and they are at lower Galactic latitudes due to the Sun's offset
from the Galactic center.

\section {Discussion} 

We have demonstrated that there are populations of stars, with distances
inferred from their turnoff magnitudes of 11 kpc from the Sun (18 kpc
from the Galactic Center), that have velocity dispersions under 30
$\rm km\>s^{-1}$.  The stars we measured, between 24$^\circ$ and 30$^\circ$
Galactic latitude, are five kpc above the Galactic plane.  We have
also observed a similar overdensity five kpc below the plane near
the anticenter.  Similar structure has been seen at every position
the SDSS has observed below 30$^\circ$ Galactic latitude.

From the
thickness of the main sequence, Paper II derived an upper limit
on the radial thickness of the structure of 6 kpc as measured from 
the magnitude range at the turnoff.  From Figure
13 of the same paper, it is shown that further down the main sequence
the FWHM is 0.7 magnitudes or less.  This results in an estimated
thickness of the structure of 4 kpc or less.  The actual thickness could
be significantly smaller than this.  The photometric errors in the colors
of the stars could account for the entire observed width of the main sequence.

The conclusion that these observations describe a ring of stars in the plane
of the Milky Way could be avoided if the distance to these stars has been
grossly overestimated from the blue turnoff stars.  If 
the absolute magnitude of these stars is not $M_g= +4.2$, but is rather
$M_g = +8$, then the stars would be much closer to the Sun.  At a distance
of 2 kpc they could be part of a known spiral arm of
the Milky Way \citep{q02}.  Similarly, if there were four magnitudes of extinction
more than we expect, then dust would be confusing our distance 
estimates.  We believe neither of these to be the case for the following reasons:

1) We know of no known star population which has the spectrum of an F star,
as seen in the numerous spectra of plate 797 (see Figure \ref{samplespecs}), 
yet has an absolute magnitude of $M_g >> 5$.  

2) The dust, if present, would redden the stars much more than is seen.
The turnoff of these stars is bluer than the thick disk 
and if unrecognized dust were present, the stars would be much
redder than $g-r = 0.3$ at turnoff.

Thus, we are left with the result that there is a ring of stars extending over
at least fifty degrees of Galactic latitude near the anticenter.  The detection
of similar structure by \citet{ietal2003} both above and below the plane near  
$(l, b) = (150^\circ, +20^\circ)$ and $(l, b)  = (123^\circ, -19^\circ)$ 
increases the range of Galactic latitude over which the structure 
extends to $100^\circ$ ($53^\circ$, as seen from the Galactic Center).  

One may estimate the metallicity of the stars in the ring from the plate 797
spectra.  We use the combination of $(g-r)_0$ color and 
Ca II K $\lambda$ 3933\AA\ equivalent width to separate effective temperature
and metallicity.  Following Figure 2 of \cite{wbg99}, we find
that objects with $0.2 < (g-r)_0 < 0.3$, have an average W(K) = 4.4\AA\ ,
which corresponds to an average $T_{eff} \sim 6400 \>\rm K$ and a typical
metallicity of $\rm [Fe/H] = -1.6$.  The scatter is
large on this metallicity (about $\pm 0.3$).  The derived metallicity, which must be considered preliminary, seems reasonable, since the turnoff is 
bluer than thick disk stars (at brighter $g$ magnitude), implying that
the structure has lower metallicity (or is younger).
The turnoff color of the stream is the same as the stars towards the center
of the Galaxy (Paper II), which we believe are spheroid stars.  Since
the Ca II K metallicities and turnoff colors are similar to the spheroid
stars, metallicities and ages similar to spheroid stars is implied.

Using star counts, we explore the extent of the ringlike structure in
more detail.  In addition to the SDSS data on the Celestial Equator
that was analyzed in detail in Paper II, the SDSS has obtained
photometric data below 30$^\circ$ Galactic latitude from Galactic
longitude 180$^\circ$ to 200$^\circ$ (SDSS stripes 30-37).  We measured
the turnoff color and magnitude of the structure in 23 pieces of sky
in this general direction.  Each patch of sky was $5^\circ \times 2.5^\circ$
and oriented along an SDSS stripe.  The turnoff magnitude was determined
from the peak of a plot of constant color derived from a Hess diagram such
as Figure \ref{cmd30-34}, and the turnoff color was determined from the peak of a plot
of the number of stars versus $(g-r)_0$ color at the turnoff magnitude.
The turnoff is at about $(g-r)_0 = 0.27$, with a small scatter, in all 
measured directions.  We measured the turnoff color and magnitude in
a similar region near plate 797 in the same way.  The turnoff color in this
direction was slightly redder, probably because of improper reddening
correction in this direction (see Paper II),
which has E(B-V) $\sim 0.08$, vs. E(B-V) $\sim 0.04$ 
for the other directions listed in Table 2 \citep{sfd98}.  

For each of these 24 datasets, we counted the number of turnoff stars. 
These stars were within 0.01 mag of the turnoff color in $(g-r)_0$ and
within 0.275 magnitudes of the turnoff magnitude in $g_0$.  The log of
the number counts of turnoff stars vs. $Z = d \sin(b)$, where $d$ is the
distance to the ring as implied from the turnoff color and an assumed absolute
magnitude of $g_0 = 4.2$ for these stars, is shown in Figure \ref{pltzb}.
We find a reasonable fit to an exponential drop off
in stellar density as function of $|Z|$ at a distance 
of 18 kpc from the Galactic center.  The
scale height fit to Figure \ref{pltzb} is $h_Z = 1.6\pm 0.5\rm \> kpc$, though
it depends on separation of stream stars from background ``spheroidal halo''
stars; $h_Z = 3$ kpc is an upper limit.  The
blue turnoff stars disappear rapidly further than about
5 kpc above or below the Galactic plane in Z in all directions we have
examined. 

Five different types of symbols are used in Figure \ref{pltzb} to separate
the stars counts into Galactic longitude bins.  Though there is quite a bit of
scatter in the diagram, it appears that at a fixed height above the disk, the
number counts are higher at $l \sim 225^\circ$ than they are at 
$l\sim 185^\circ$ by about 25\%.
This effect may be simply be due to the geometry, as the fraction of the ring
intercepted in each direction is different.
At $l\sim 185^\circ$, the distance to the stream (10.7 kpc) is about 12\%
less than that at $l\sim 225^\circ$ (12 kpc, converting the turnoff 
magnitudes in Table 2 to approximate distances).  Thus, the solid angle
intercepted is about 25\% less for the lower longitude regions, roughly 
consistent with the counts plotted in Figure \ref{pltzb}.  This again argues
that the structure we are seeing is coherent with relatively
uniform density, at least for the northern Galactic latitudes.  
The planar projected distance $R = \sqrt(X^2+Y^2)$ of the 
pieces of the structure listed in Table 2 are 19.5, 17.2, 17.8, and 17.6 
kpc. The first quoted distance, to the region of Plate 797, is significantly farther
than any of the other measured distances from the Galactic Center. 
Its turnoff color quoted in Table 2 and in Paper II is 
slightly bluer than that of the other structures, implying that reddening
is likely to be overestimated towards this direction on the sky. 
Correction of this error will only increase the distance disparity
between northern and southern components of the structure.

The similarity of this scale height to the
proposed scale height for a metal weak thick disk
\citep{mff90,m93,bdrcrfnh2002}, and the embarrassingly good fit to the
star number counts in Figure 20 of Paper II, made it difficult to claim
that we had found a ring of stars in Paper II, even though we had observed
a narrower main sequence than expected from an exponential disk.  The
additional evidence in the form of a narrow velocity dispersion strengthens
the ring identification; the structure
has a much smaller velocity dispersion than is expected for a large
scale-height exponential disk.

If we identified all of the stars in this structure with an exponential disk
of large scale height, then we would have been forced to include all of the
stars observed towards the Galactic center in Paper II in the same disk
structure, leaving no stars left in the power law spheroid distribution.  That
consequence of the exponential disk conclusion would have been difficult
to explain, given the long history of measurements of the spheroid component.
The spectra we obtained in this paper show a strong power law spheroid
component near the ringlike structure, with the same density and radial 
velocities as standard Galactic models predict.

Still, the
possibility that the stars in this Monoceros structure are related to
those solar neighborhood stars called metal weak thick disk stars 
\citep{n94} is intriguing.  It is interesting that the inferred scale height,
metallicity, and Galactic rotation speed of this population is in good
agreement with those of the proposed metal-weak thick disk.

It is also interesting to compare the structure seen in
the outer parts of our Galaxy to shells of stars seen in the
outskirts of some elliptical galaxies.  The radial narrowness of the
structure ($< 4\rm\> kpc$), the
tightness of the velocity dispersion, and the Z extent (5 kpc above and
below the plane), as well as the uniform stellar population (color of
turnoff and metallicity from CaII K line equivalent width) all are
reminiscent of shells such as those seen in \citet{mc83}.  If the
structure is a shell, it may be possible to use the narrow velocity
dispersion and Z extent to give information about an early Galactic
collision with another massive object \citep{t99} and to constrain the
dark halo potential more precisely \citep{mk98}.

One is, of course, curious as to the total mass of the structure.
We estimate lower and upper bounds on the stellar mass in the structure
as follows:

A lower bound follows if the structure is two streams, and extends  only as
far as we see it, over an arc of $50^\circ$ (rather than extending 
all the way around the Milky Way).   
In Figure \ref{pltzb} we fit an exponential profile to the $Z$ extent 
for $15^\circ < |b| < 30^\circ$.   The minimal star 
count assumes that the maximum star count (the middle of the stream) is 
at exactly $|b| = 15^\circ$ so there are twice as many stars per 
stream as there are over the range of Galactic 
latitudes $15^\circ < |b| < 30^\circ$.
In this case, we integrate the total number of F turnoff stars 
per square degree (as seen from the Galactic center), based on 
Figure \ref{pltzb} in this fashion:

\[ N_{\rm Monoceros\> turnoff\> stars} =  \frac{180^\circ}{\pi} \frac{18 \rm kpc}{10 \rm kpc} \times 50^\circ \times \frac{10000}{12.5 \rm deg^2} \times 2 \times 2 \times \int_{15^\circ}^{30^\circ} e^{-d \sin|b|/h_Z} db \]

Where $180^\circ/\pi$ handles the degrees to radians conversion; 
 18 kpc/10 kpc converts an angle subtended at the sun to 
one subtended from the Galactic center; $50^\circ$ is the arc over 
which we actually observe Monoceros structure
stars; $10000/12.5\rm deg^2$ are area and normalization factors; the first 
factor of two is for two streams, one above and one below the Galactic plane; the second factor of 2 counts the stars
with $|b| < 15^\circ$; and $d \sin|b| = |Z|$, 
where $d = 10 $ kpc and $h_Z = 1.6$ kpc toward $l \sim 180^\circ$.  
This yields $N_{MTS} = 3.6\times10^5 $ turnoff stars.  We estimate from 
Paper I, Paper II and \citet{oetal01} the ratio of turnoff stars in the 
magnitude-color box of Figure \ref{pltzb} to total solar masses in stars in a 
globular cluster like Palomar 5 to be about 1:50.  Thus the lower 
limit on stellar mass in the Monoceros structure 
is $2\times 10^7 M_\odot$.    We note that this
is considerably larger than our lower limit mass from Paper II, and is
based on the fact that we now have SDSS photometric data to lower $|b|$ and
observe the number counts that continue to rise as $|b|$ decreases. 

We follow a similar prescription in placing an upper limit, changing
the $50^\circ$ of arc to a full $360^\circ$ and integrating from
$0^\circ < |b| < 30^\circ$ in the above equation (removing one 
factor of 2).  This yields  $N_{MTS} = 9 \times 10^6$ and a total 
mass in stars of $5\times 10^8 M_\odot$.  If many stars have evaporated
from the structure, or the stars are strung out over multiple orbits of
the Galactic Center, then the initial stellar mass of a progenitor
galaxy could be larger.

The implied rotation velocity of the stars in this structure of 110 
$\rm km\>s^{-1}$ is very close to the rotation velocity of the stars 
that \cite{gwn2002} attribute to the remnant of a Galactic merger which
may have puffed up the thick disk, implying that the stars in our structure and the 
stars identified in \citet{gwn2002} could have a common origin.  If this
is the case, then the merging galaxy likely had a stellar mass substantially
larger than $10^9 M_\odot$; since the \citet{gwn2002} stars are located
at 3-4 kpc from the Sun and the stars we identified are 9 kpc or more
from the solar position, a common origin for the stars would imply that
the stars were spread through a substantial fraction of the Galactic disk.




\section{Conclusions}

	We find a ring of stars in the plane of the Milky Way, at a distance of 18 kpc
from the Galactic Center.  We have traced it from $l=180^\circ$ to $l=227^\circ$, and it may
completely encircle the Galaxy at low latitudes ($|b| < 30^\circ$).  The structure
extends 5 kpc above and below the plane of the Galaxy.  The stars at negative Galactic latitude
are about 2 kpc further from the Galactic center than those above the Galactic equator.
The radial thickness of the ring is less than 4 kpc FWHM.  Without more data in the plane,
it is difficult to determine a scale height for the stars, but we constrain $h_Z < 3 $ kpc.
The stellar density at a given height above the plane is constant over the range of Galactic
latitudes we have surveyed.  Given that there is an asymmetry between the distance to the northern
and southern portions of the structure, it is possible that there is additional substructure yet
to be discovered.

	The excess stars discovered at low latitudes in Paper II, and interpreted in this paper
as a ring of stars, are not associated with the thin disk, thick disk, proposed metal-weak thick
disk, or power law spheroid star distributions.  The stars are not part of the
Sagittarius dwarf tidal stream.  The evidence for this includes:

(1) At distances of 5 kpc above the Galactic plane, and 18 kpc from the Galactic Center, there are
no thin disk stars.  Likewise, there are two orders of magnitude fewer thick disk stars than power law spheroid stars at these distances and directions.

(2) From the spectra of 327 F stars on special plate 797, we 
find a mean radial velocity of 74 $\rm km\>s^{-1}$ and velocity dispersion of 27 $\rm km\>s^{-1}$ for stars 
at $(l,b) = (198^\circ, -27^\circ)$. 
The velocity dispersion of the low latitude structure in three additional
directions: $(l, b) = (225^\circ, 28^\circ)$, $(182^\circ, 27.5^\circ)$, and $(188^\circ, 23.5^\circ)$, are 22 $\rm km\>s^{-1}$, 30 $\rm km\>s^{-1}$,
and 30 $\rm km\>s^{-1}$, respectively.  Since the spectra are obtained over several degrees of sky, over
which the mean radial velocity of the structure can shift, the
intrinsic dispersion of the stars in the ring could be slightly smaller.
These velocity dispersions are much smaller than expected from
the spheroid (120 $\rm km\>s^{-1}$) or thick disk (55 $\rm km\>s^{-1}$) populations.  
One would expect that the proposed metal-weak thick disk with large scale height
would have an even larger velocity dispersion than the thick disk, which does not match the observations.  The combination of large numbers of stars at 2 -- 5
kpc above the plane and a narrow velocity dispersion makes it difficult to identify
these stars with any known or proposed exponential disk distribution.
No simple warp or flare of an exponential disk can simultaneously explain the excess stars above the plane and
below the plane, with a velocity dispersion under 30 $\rm km\>s^{-1}$.

(3) The narrow width of the main sequence in Hess diagrams such as Figure 7 (or Figure 12 of Paper II) is inconsistent with exponential disk models of 
any scale height/length, even if photometric errors are assumed to be negligible (see Figure
13 of Paper II).  The photometric errors alone can explain the entire observed width of the main
sequence in this diagram.

(4) The stars in Sp198-27-19.8 have Ca II 3934\AA\ equivalent widths and
$(g-r)_0$ turnoff colors which suggest metallicities of 
$[Fe/H] = -1.6 \pm 0.3$, consistent with stars in the halo or a metal
weak thick disk, and inconsistent with those of a thin or thick disk
warp or flare and also inconsistent with those of the more metal poor
and bluer Sagittarius stream stars.  The turnoff color is consistent with that
of spheroid stars. 

(5) The alignment of the density structure of the ring with the Galactic plane is inconsistent with the orbit of the Sagittarius dwarf tidal stream (which
follows a nearly polar orbit).

We measure the radial velocity and velocity dispersion of the tidal stream of the Sagittarius
dwarf galaxy at $(l, b) = (165^\circ, -55^\circ)$ to be -111 $\pm 5 \rm \>km\>s^{-1}$ with a dispersion of 20 $\rm km\>s^{-1}$.  The dispersion of the Sagittarius stream is narrower than
our measured dispersion for the ring of stars.  If we interpret the ring of stars as
the result of the tidal disruption of a smaller galaxy by the Milky Way, then this
leads us to surmise that the mass of the progenitor was significantly larger than the
Sagittarius dwarf galaxy.  

We have estimated that there are between $3.6 \times 10^5$ and  $9 \times 10^6$ turnoff
stars in the ring, depending on the assumptions as to how far around the Galaxy it goes,
and what its density profile is in the plane of the Milky Way.  We estimated from the
globular cluster Palomar 5 that the stellar mass of the structure is fifty times larger
than the number of turnoff stars.  Therefore, the inferred mass in stars of the ring
is between $2 \times 10^7$ and $5 \times 10^8 \rm M_\odot$.  If there is a significant
fraction of dark matter associated with the stars, the mass could be ten times larger.
The stellar mass of the
Sagittarius dwarf galaxy ($1.8 \times 10^7 \rm M_\odot$, assuming a mass to light 
ratio of 1) is at the lower limit of our calculated mass range for the ring.  This is
consistent with our conjecture that the progenitor would be more massive than the
Sagittarius dwarf from the larger measured velocity dispersion.

The radial velocities of four pieces of the ring are tabulated in Table 2.  Using the
simplistic assumption that the stars in the ring move coherently (on average) in a circular orbit,
we calculated a prograde rotation of the ring at a speed of $110 \pm 25 \rm \>km\>\>s^{-1}$.
Only the first two radial velocity determinations (Sp198-27-19.8 and Sp225+28-19.6)
were significant in this calculation, since the other two directions were very close
to $l = 180^\circ$.  It is interesting that the two measurements disagree at the
2.4 $\sigma$ level, indicating that the kinematic and spatial structure of the
ring is more complex.  This is supported
by the disparate Galactocentric distances measured to these two portions of the structure.

This feature of Galactic structure seen in the constellation Monoceros, confirmed here 
by both photometric and kinematic techniques shows the depth and accuracy of new
wide area surveys such as the SDSS and provides a new set of
tracers for exploring the structure and evolution of our Milky Way.




\acknowledgments


We thank Rich Kron for useful discussions.  HJN acknowledges funding from 
Research Corporation. We thank the anonymous referee for suggestions
which improved the paper.

Funding for the creation and distribution of the SDSS Archive has been
provided by the Alfred P. Sloan Foundation, the Participating
Institutions, the National Aeronautics and Space Administration, the
National Science Foundation, the U.S. Department of Energy, the
Japanese Monbukagakusho, and the Max Planck Society.  The SDSS Web
site is http://www.sdss.org/.

The SDSS is managed by the Astrophysical Research Consortium (ARC) for the 
Participating Institutions. The Participating Institutions are The University 
of Chicago, Fermilab, the Institute for Advanced Study, the Japan Participation 
Group, The Johns Hopkins University, Los Alamos National Laboratory, the 
Max-Planck-Institute for Astronomy (MPIA), the Max-Planck-Institute for 
Astrophysics (MPA), New Mexico State University, University of Pittsburgh, 
Princeton University, the United States Naval Observatory, and 
the University of Washington.

\clearpage

\figcaption { 
	A $g_0, (g-r)_0$ color magnitude diagram of SDSS stars near
	$(l,b) = (198^\circ, -27^\circ)$.  The Milky Way's thick disk turnoff is
	seen as a wide structure near $(g-r)_0 = 0.3, g_0 = 19$.  Note,
	however, the additional faint turnoff at $(g-r)_0 = 0.2,\>
	g_0 = 19.5$. This is where the spectroscopic targets for plate 797
	observations were selected (triangles).
\label{selectedstars}}

\figcaption { Histogram of radial velocities for 327 blue stars with
        $19.1 < g_0 < 20.3$ and $0.158 < (g-r)_0 < 0.3$ in the direction
        $(l,b) = (198^\circ, -27^\circ)$.  The stars have an average 
	heliocentric radial velocity of $\rm 74 \>\>km\>s^{-1}$ with a 
	remarkably small 1D velocity dispersion of $\rm \sigma = 27
        \>km\>s^{-1}$ after subtraction in quadrature of typical instrumental
	errors of $\rm 20 \> km \> s^{-1}$. 
        The distance to the stars
        was calculated from the turnoff magnitude listed in Table 2, assuming
        an absolute magnitude of $M_g = 4.2$ for turnoff stars.
	Also plotted are several models, representing
        expected contributions and projected radial velocities of
        stars from the Milky Way's thin disk (red) and thick disk
        (green), both negligible, for objects of this color at this
        distance from the Galactic center, and for the stellar
        spheroidal halo (blue).  The black dotted line is the sum of
        the thin disk, thick disk and halo components and the thin
        solid line represents a Gaussian fit to the `extra' stars,
        where the contribution of the halo has been adjusted to
        minimize the overall fit (thick black line).  The density excess
	and narrow dispersion are striking for these stars, which 
	are apparently 20 kpc from the Galactic center.
\label{rvhist}}

\figcaption { 
	The difference of SDSS radial velocities and catalog
	velocities \citep{mlgg86} of bright blue stars in the open cluster M67
	vs. the fiber number on SDSS plate 321.  Each plate has 640
	fibers, divided into two (left and right) halves of 320 fibers
	each.  The wavelength calibrations are determined separately
	for the left and right halves based on arc lamp exposures
	before and after the target exposures.  There is a systematic
	difference in the SDSS velocities on the left and right halves
	of the plate of amplitude $4 \rm \>km\>s^{-1}$; this is an
	estimate of the absolute error in radial velocities determined
	throughout this paper.  The one sigma dispersion of $11\rm
	\>km\>s^{-1}$ is a measure of the rms scatter for
	individual measurements of bright star radial velocities.  For
	lower S/N observations, such as those in plate 797, we estimate a one
	sigma dispersion of $20\rm \>km\>s^{-1}$.
\label{stdplot}}

\figcaption{
	Histogram of errors in the determination of radial velocities for
	all SDSS spectra from many plates in the color-magnitude box of:
	$0.15 < (g-r)_0 < 0.36; 18.9 < g_0 < 20.0$.  These are differences
	in the automated radial velocities determined for multiple
	SDSS observations of the same star on different spectroscopic 
	plates.  There are N=146 pairs, with $\sigma/\sqrt{2} = 16.3\rm \> km\>s^{-1}$,
	which gives us a good estimate of our radial velocity accuracy
	for objects with typical S/N=10.
\label{rverrplt}}

\figcaption{
	The color magnitude (``Hess'') diagram of SDSS stars in the region of the Sagittarius
	stream near $(l,b) = (165^\circ, -55^\circ)$.  The greyscale in each color-magnitude pixel is proportional
	to the square root of the number of stars in that bin.  The black box indicates the area where
	horizontal branch and blue straggler stars from the Sagittarius stream are found. The white box matches the selection box used at lower
	Galactic latitudes and represents a high-latitude control field for determining
	the extent of the low latitude blue stream stars.
	Objects in the figure were selected with the color cuts:
	$0.4 < (u-g)_0 < 1.8,  -1 < (g-r)_0 < 1.5$.
\label{cmdsagbox}}

\figcaption {
	Histogram of radial velocities for very blue horizontal
	branch and blue straggler stars 
	in the direction of the Sag South stream (black box of Figure \ref{cmdsagbox}).
	The stars have an average 
	heliocentric radial velocity of $\rm -111 \>km\>s^{-1}$ and 
	a velocity dispersion of $\rm \sigma \sim 20 \>km\>s^{-1}$, (after removing an
	instrumental spread of $\rm 7 \> km \> s^{-1}$).  consistent 
	with models of the Sagittarius South tidal stream.
\label{rvsagsouth}}

\figcaption{
	The color magnitude density diagram of SDSS imaging (all stellar objects) in the a region surrounding the areas Sp182+27-19.4 and Sp188+24-19.3 centered on $(l,b) = (185^\circ, 25^\circ)$.  The greyscale in each color-magnitude 
pixel is proportional to the square root of number of stars in that bin.  The white box indicates the selection area 
	used for picking SDSS spectra of stars in the faint blue turnoff.  
	Objects in the whole figure were selected with the color cuts:
	$0.4 < (u-g)_0 < 1.8,  -1 < (g-r)_0 < 1.5$.  
\label{cmd30-34}}

\figcaption {
	Five SDSS spectra plotted in the wavelength range 3850--5000\AA\ 
	(the full SDSS spectra extend over the range 3780--9100\AA\ ).
	All spectra are offset and scaled to fit on the page, with a zero
	flux level drawn under each and a vertical reference line at H-beta.
	Top to bottom: A G star in the M67 standard plate 321 field;
	A BHB star in the Sagittarius dwarf southern tidal stream; 
	An F turnoff star in the Sp225+28-19.6 area of 
	Monoceros; and two plate 797 stars representing the bright 
	and faint ends of the set of F turnoff stars observed spectroscopically.
\label{samplespecs}}

\figcaption {
	Histograms of radial velocities, with Galactic model fits, 
	for the spectra of stars of the last four entries of Table 2.
        The distance to the stars
        was calculated from the turnoff magnitude listed in Table 2, assuming
        an absolute magnitude of $M_g = 4.2$ for turnoff stars.
	The central velocities and velocity dispersions of three low latitude
	panels are consistent with one tidal stream circling the Galaxy 
	at a distance of 18 kpc from the Galactic center.  The dispersion
	of the stream is consistent with the 27 $\rm km\>s^{-1}$ seen
	in Figure \ref{rvhist}.  The lower right panel shows that the
	low latitude structure seen in the other panels is not present at
	Galactic latitude $b \sim -55^\circ$.
\label{rvhists}}

\figcaption {
	Galactic (X,Y,Z) plane projection plots of one possible 
	orbit for stars in the stream.  This sample model has parameters listed
	in Table 3.  The Galactic center and Sun at 
	(X,Y,Z) = (-8,0,0) kpc are marked.  The four spots at $\rm R\sim 18 \>kpc$
	from the G.C. are where the SDSS has photometric and kinematic
	data showing a faint blue turnoff.  The radial velocity of the
	sample orbit matches that observed at the four observed regions
	to within about 7 $\rm km \> s^{-1}$ (see Table 3). The model has 
	a flattened logarithmic potential with potential flattening 
	parameter of $q_p = 0.69$.  
\label{pltorbit}}

\figcaption { 
	Equatorial ($\alpha_{2000},\delta_{2000}$) Aitoff projection plot of one
	possible orbit of the stream (from Table 3).  The stars are all 
	at about 18 kpc from the Galactic center, and are 
	nearly all at $|b| < 30^\circ$ (bounded by red lines).  The color of the plotted orbit
	changes as noted in the figure legend to represent projected 
	heliocentric radial velocity for stars observed from the Sun.  The
	four black points are the positions of the observed SDSS spectroscopic
	and photometric data. The black points are offset a bit from the stream centers
	to reflect the fact that the data do not probe to $|b| < 15^\circ$.
	A shift of about about
	1 kpc is assumed.  The stripes of the SDSS survey area (as planned),
	are plotted in magenta.  Only every other stripe in the Northern Galactic
	hemisphere is plotted.  
\label{radec}}

\figcaption{
	Height above (or below) the plane $|Z|$ vs. turnoff 
	faint blue star counts in a color magnitude box for stars near
	the turnoff if $12.5 \rm \>deg^2$ boxes in 
	the low latitude stream. 
	Though the estimate of background (smooth stellar halo stars) is
	uncertain,  one can fit an exponential profile to the density of
	stars in the stream vs. height above the disk in the range where
	data is available.  This fit yields a formal scale height of
	$h_Z = 1.6\pm 0.5 $ kpc for this choice of constant (700) 
	background counts.
	The data points are broken into Galactic longitude ranges.
\label{pltzb}}

\clearpage
\begin{deluxetable}{rrcccccrrcr}
\tabletypesize{\scriptsize}
\tablecolumns{11}
\footnotesize
\tablecaption{Blue stars near (l,b) = (198, -27) (Table Stub only)}
\tablewidth{0pt}
\tablehead{
\colhead{R.A.} &
\colhead{Dec.} &
\colhead{SDSS ID} & \colhead{Fiber ID} &
\colhead{$g_0$} & \colhead{$(g-r)_0$} &
\colhead{$(u-g)_0$} & \colhead{RV} & \colhead{W$_K$} &
\colhead{color flag} & \colhead{E(B-V)}\\
\colhead{$^\circ$} & \colhead{$^\circ$} & \colhead{r-re-c-f-id} & \colhead{plate-mjd-fiber} & \colhead{mag} & \colhead{mag} & \colhead{mag} & \colhead{km s$^{-1}$} & \colhead{\AA } & \colhead{} & \colhead{mag} }

\startdata

 70.598290 &  -0.213427 & 0125-7-3-546-0318 & 797-52263-318 & 19.632 &  0.232 &  0.984 &   30.1 &  6.19 & 1  & 0.0606\\
 70.603553 &  -0.237871 & 0125-7-3-546-0219 & 797-52263-313 & 19.919 &  0.280 &  0.731 &   62.9 &  3.36 & 1  & 0.0573 \\
 70.622982 &   0.047201 & 0125-7-4-546-0262 & 797-52263-359 & 19.788 &  0.264 &  0.708 &   89.0 &  5.01 & 1  & 0.0768 \\
 70.625600 &  -0.057840 & 1752-0-3-332-0329 & 797-52263-358 & 19.230 &  0.272 &  1.042 &   78.4 &  4.11 & 1  & 0.0798 \\
 70.652157 &  -0.544781 & 1752-0-2-333-0301 & 797-52263-311 & 19.397 &  0.223 &  0.892 &  145.5 &  3.76 & 1  & 0.0346 \\
 70.788270 &   0.555787 & 0125-7-5-547-0374 & 797-52263-339 & 19.229 &  0.138 &  1.122 &   77.0 &  5.03 & 0  & 0.0788 \\
\enddata
\end{deluxetable}

\begin{deluxetable}{lrrrrrrrrr}
\tablecolumns{10}
\footnotesize
\tablecaption{Summary of Stream detection pieces}
\tablewidth{0pt}
\tablehead{
\colhead{Name} & \colhead{$\alpha$} & \colhead{$\delta$} & \colhead{l} & \colhead{b} &
\colhead{$<v_R>$} & \colhead{$\sigma(v_R)$} & \colhead{$g_0$\tablenotemark{1}} & 
\colhead{$(g-r)_0$\tablenotemark{1}}& \colhead{$v_{circ}$} }

\startdata
Sp198-27-19.8\tablenotemark{2}& 72 & 0& 198 & -27 & 74 $\pm 5$ & 27 & 19.8 & 0.24 & -5$\pm 54$  \\
Sp225+28-19.6& 133 & 2& 225 & 28 & 103 $\pm 6$ & 23 & 19.6 & 0.26 & 139$\pm 27$\\
Sp182+27-19.4& 117 & 38& 182 & 27 & 22 $\pm 8$ & 30 & 19.4 & 0.28 & ---\\
Sp188+24-19.3& 115 & 32 & 188 & 24 & 34 $\pm 7$ & 30 & 19.3 & 0.28 & 12$\pm 200 $ \\
\hline
Sag. South Strm& 35 & 0&165 & -55 & -111 $\pm 5$\tablenotemark{3} & 20\tablenotemark{3} & 21.5\tablenotemark{3} & 0.22\tablenotemark{3} & --- \\

\enddata
\tablenotetext{1}{mag and color of faint blue Monoceros turnoff except where indicated}
\tablenotetext{2}{Plate 797}
\tablenotetext{3}{For the Sagittarius stream, this $<v_R>$ and $\sigma$ refer to Sagittarius stream blue horizontal branch and blue straggler stars rather than those in the Monoceros structures. The color and mag are those of the Sag. stream.}
\end{deluxetable}

\clearpage

\begin{deluxetable}{lr}
\tablecolumns{2}
\footnotesize
\tablecaption{Stream Model parameters}
\tablewidth{0pt}
\tablehead{
\colhead{Parameter} & \colhead{Value} }

\startdata
 
Halo Potential model 1 & ln(s) \\
Potential Flattening for model 1 & $q_p = 0.69$ \\
Halo Potential model 2 & Monet, Richstone, \& Schechter \\
Potential Flattening for model 2 & $q = 0.78$ \\
$(x_0,y_0,z_0)$ & $(-15.498, -7.487, 4.633) \rm \>kpc$ \\
$(vx_0,vy_0,vz_0)$ & $(-83.6, 202.4, 59.4) \>\rm km\>s^{-1}$\\
Period & 600 Myr\\
Semi-Major Axis of Orbit & 18 kpc \\
Predicted (Observed) RV at Sp225+28-19.6 & 96 (103) $\>\rm km\>s^{-1}$\\
Predicted (Observed) RV at Sp198-27-19.8  & 72 (74) $\>\rm km\>s^{-1}$\\
Predicted (Observed) RV at Sp182+27-19.4 & 25 (22) $\>\rm km\>s^{-1}$\\
Predicted (Observed) RV at Sp188+24-19.3 & 36 (34) $\>\rm km\>s^{-1}$\\
Closest approach of orbit to Sp198-27-19.8 & 0.6 kpc \tablenotemark{1}\\
Closest approach of orbit to Sp182+27-19.4 & 0.7 kpc \tablenotemark{1}\\
Closest approach of orbit to Sp188+24-19.3 & 0.3 kpc \tablenotemark{1}\\

\enddata
\tablenotetext{1}{The closest approach is actually to a position 1 kpc lower
in $|Z|$ to that of the observed stars.}

\end{deluxetable}

\clearpage

\setcounter{page}{0}

\plotone{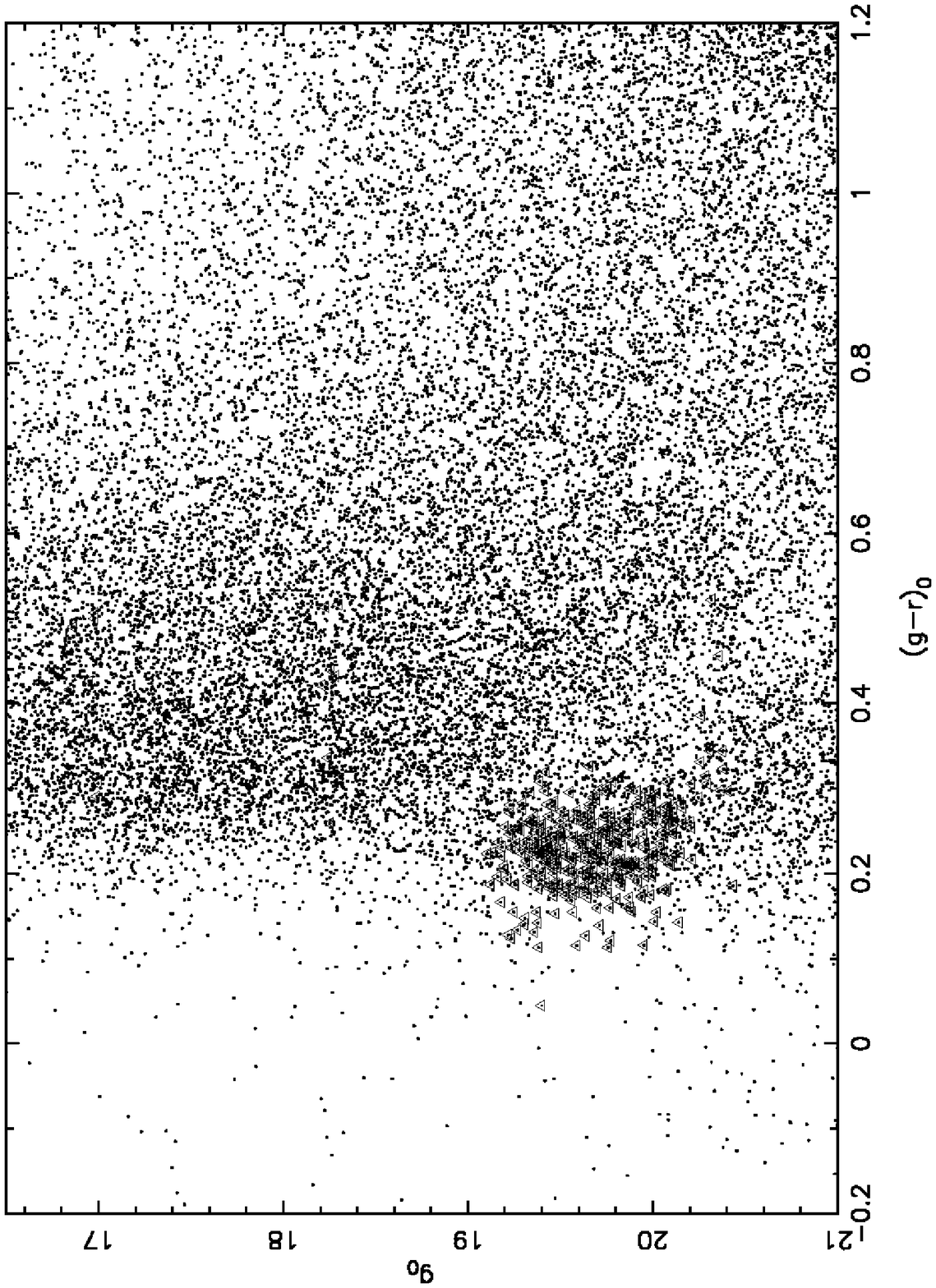}

\plotone{f2.eps}

\plotone{f3.eps}

\plotone{f4.eps}

\plotone{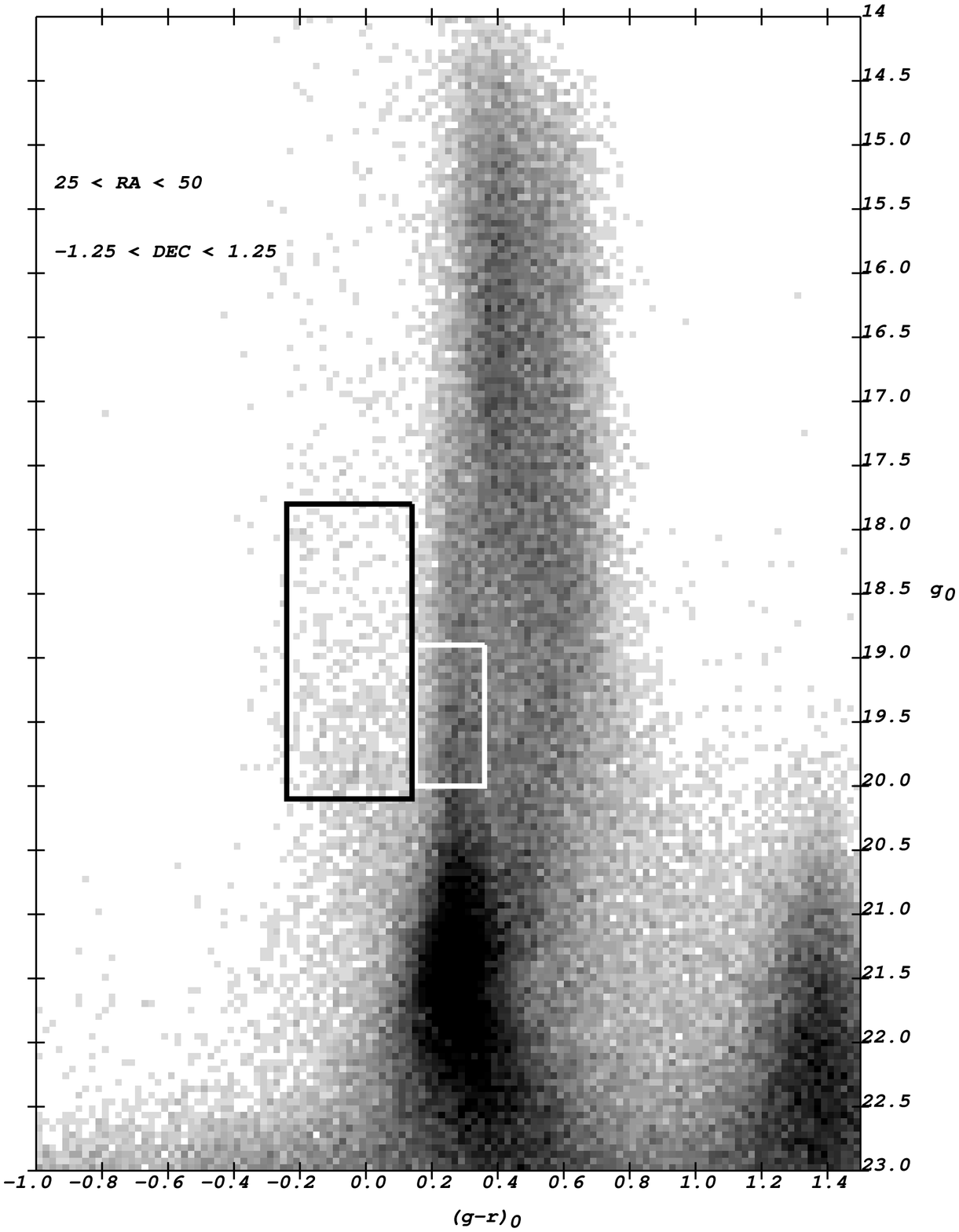}

\plotone{f6.eps}

\plotone{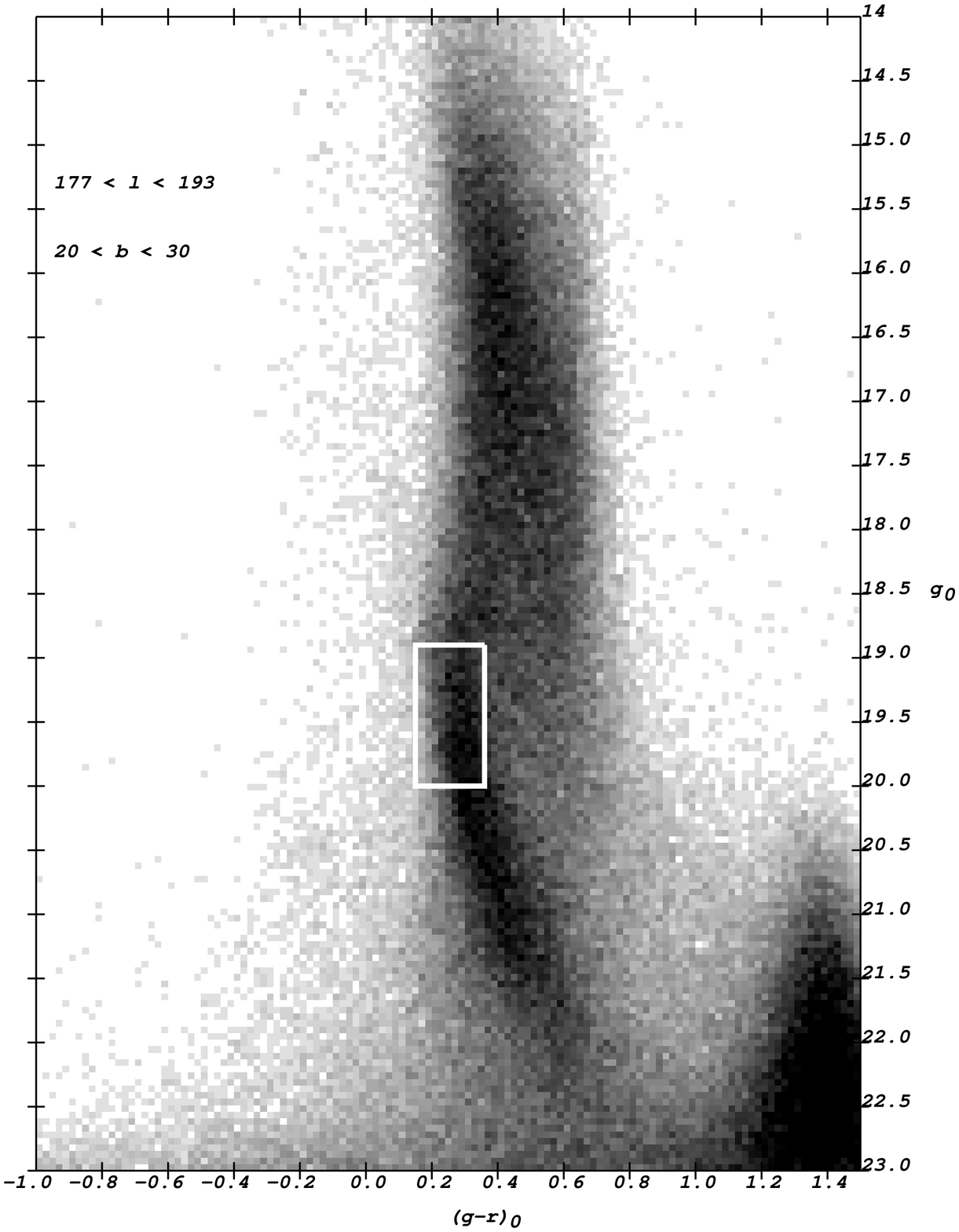}

\plotone{f8.eps}

\plotone{f9.eps}

\plotone{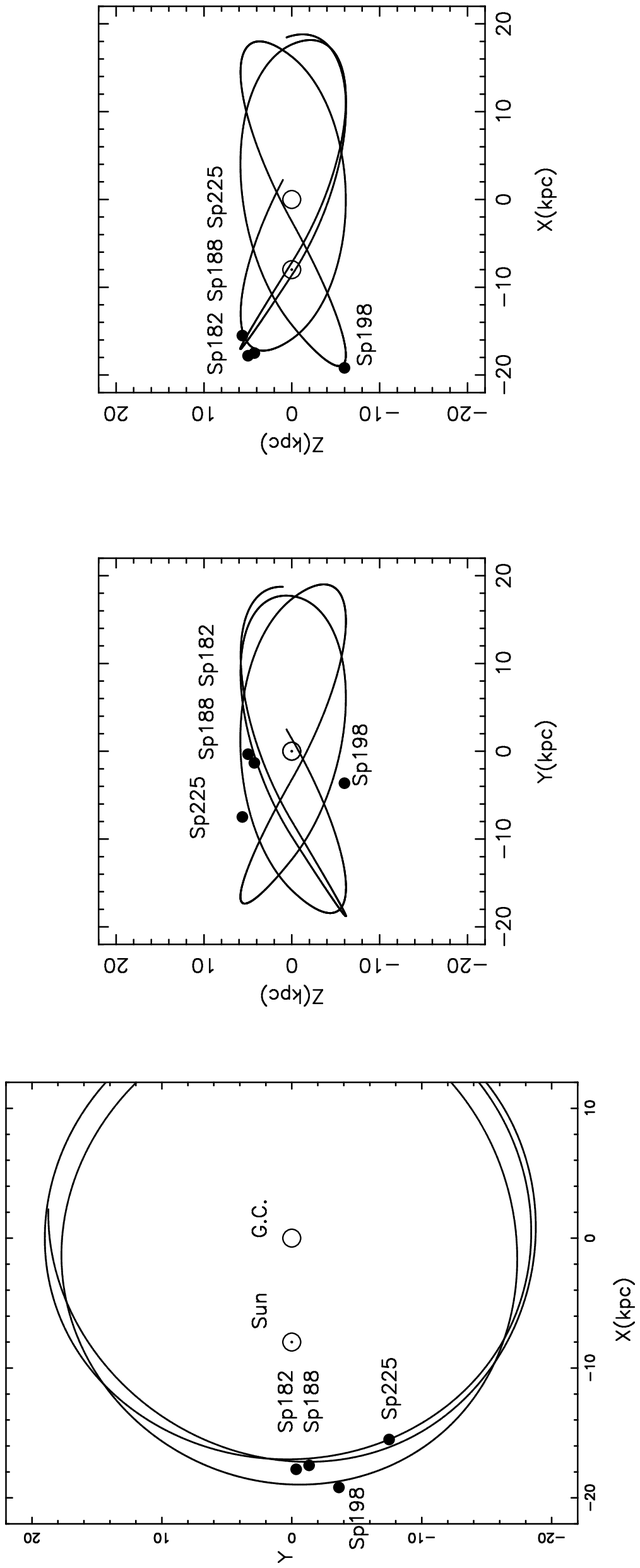}

\plotone{f11.eps}

\plotone{f12.eps}

\clearpage

\setcounter{page}{1}


%

\setcounter{section}{0}

\flushleft

\section{Erratum: A Low Latitude Halo Stream around the Milky Way}


The zeropoints of the stellar templates used to measure radial velocity in the main body of this paper
have been found to be systematically in error.  Correction of the radial velocities
significantly increases the derived circular velocity of the stars in the planar stream
to $215 \pm 25 \rm \> km\>s^{-1}$.  The velocity dispersion of the stream is
somewhat lower than earlier results with the modified analysis.

Two types of stars were studied in this paper.  The original template for stars of type F, used
to study the ``Monoceros arc" Galactic structure, was incorrectly zeropointed by
-20 $\rm km\>s^{-1}$.  The original template for stars of type A, used to measure
the Sagittarius dwarf tidal stream, produced radial velocities systematically
shifted by -49 $\rm km\>s^{-1}$. 

A cross-correlation of Sloan Digital Sky Survey (SDSS) spectra with templates from the 
ELODIE survey (Soubiran, Katz and Cayrel 1998) was performed to find new radial velocities for each star
(D. Schlegel, private communication).  This showed that our radial velocities were systematically 
shifted by an amount that depends on the type of the star observed and on the original template 
against which it was cross-correlated.  

To determine the measurement error with the new templates,
we identified 445 F-type stars and 1109 A-type stars which had been observed twice 
by the SDSS.  These stars were chosen with the color and magnitude criteria used to select stars in Figures 6
and 9.  The errors in the F stars were a good match to a Gaussian with a sigma of 28 $\rm km\>s^{-1}$.  The errors in the
A star comparison were significantly non-Gaussian, with large tails.  A $\chi^2$ fit to a Gaussian (similar
to the technique we use in this paper to measure the width of the streams) yielded a sigma of 35 $\rm km\>s^{-1}$.  
Dividing by $\sqrt{2}$ to reflect two independent measurements, we derive a random error of 20 $\rm km\>s^{-1}$ for 
F stars and 25 $\rm km\>s^{-1}$ for A stars.  The template matching 
errors in these blue (type A) stars using ELODIE spectral templates are somewhat larger than the errors with our 
previous analysis, but we found it useful to use ELODIE spectral templates to ensure that the zeropoints were accurate.

Additionally, we examined the measured stellar stream dispersions.  
Electronic versions of Figure 2, Figure 6, and Figure 9 of our paper are presented here with the corrected radial velocity
determinations.  The data were selected as described in the original paper.

Table 1 has been regenerated in its entirety, replacing columns 8 and 10.  The radial velocity in column 8 has been replaced
with the radial velocity determined from cross-correlation with ELODIE templates.  The status flag in column 10 now
indicates stars which were used to generate Figure 2.  A ``0" indicates that the star was either outside of the color box
or had a high cross-correlation error, and a ``1" indicates that the star was used to fit stream properties.

Table 2 has been regenerated using the new results as well.  Column 10 has been added to indicate the estimated number of 
spectra in the stream component.  These numbers are used to compute the error in radial velocity, as described in the
original paper.  Column 11 shows the corrected circulation velocities, which are now consistent with \citet{cmrfsl03}.
Note that the velocity dispersions of the planar stream are even tighter than originally measured, strengthening the case
that the motion is coherent.  Note that the mean velocity of the Sagittarius stream in the direction $(l,b) = (165^\circ,-55^\circ)$
is -160 $\rm km\>s^{-1}$, in line with recent simulations by \citet{mdetal}.

We would like to acknowledge Steve Majewski, who initially pointed out to us that radial velocities for stars he had measured
in the halo streams were different from our radial velocities by 20 to 50 $\rm km\>s^{-1}$ \citep{cmrfsl03}.
We also acknowledge T. Beers, C. Prieto, and R. Wilhelm for an independent RV analysis, with which we could compare our
measured radial velocities.

\clearpage

{ Figure 2 caption (revised version): Histogram of radial velocities for 234 blue stars with
        $19.1 < g_0 < 20.3$ and $0.158 < (g-r)_0 < 0.3$ in the direction
        $(l,b) = (198^\circ, -27^\circ)$.  Stars with radial velocity errors (as 	
	determined from cross correlation with an ELODIE template of similar spectral type) of 
	over 20 $\rm \> km\>s^{-1}$ were rejected.  The stars have 
	an average heliocentric radial 
	velocity of $\rm 54 \>\>km\>s^{-1}$ with a
        remarkably small 1D velocity dispersion of $\rm \sigma =$ 18
        $\rm \>km\>s^{-1}$ after subtraction in quadrature of typical instrumental
        errors of 20 $\rm \> km \> s^{-1}$.
        The distance to the stars
        was calculated from the turnoff magnitude listed in Table 2, assuming
        an absolute magnitude of $M_g = 4.2$ for turnoff stars.
        Also plotted are several models, representing
        expected contributions and projected radial velocities of
        stars from the Milky Way's thin disk (red) and thick disk
        (green), both negligible, for objects of this color at this
        distance from the Galactic center, and for the stellar
        spheroidal halo (blue).  The black dotted line is the sum of
        the thin disk, thick disk and halo components and the thin
        solid line represents a Gaussian fit to the `extra' stars,
        where the contribution of the halo has been adjusted to
        minimize the overall fit (thick black line).  The density excess
        and narrow dispersion are striking for these stars, which
        are apparently 20 kpc from the Galactic center.
}

{ Figure 6 caption (revised version): Histogram of radial velocities for very blue horizontal
        branch and blue straggler stars
        in the direction of the Sag South stream (Fig. 5, {\it black box}).
	As in Figure 2, stars with large radial velocity errors (as determined 
	from cross correlation with an ELODIE template of similar spectral type) were rejected.
        The stars have an average
        heliocentric radial velocity of -160 $\rm \>km\>s^{-1}$ and
        a velocity dispersion of $\rm \sigma \sim 22 \>km\>s^{-1}$, (after removing an
        instrumental spread of $\rm 25 \> km \> s^{-1}$).
}

{ Figure 9 caption (revised version): Histograms of radial velocities, with Galactic model fits,
        for the spectra of stars of the last four entries of Table 2.
        The distance to the stars
        was calculated from the turnoff magnitude listed in Table 2, assuming
        an absolute magnitude of $M_g = 4.2$ for turnoff stars.
        The central velocities and velocity dispersions of three low-latitude
        panels are consistent with one tidal stream circling the Galaxy
        at a distance of 18 kpc from the Galactic center.
        The lower right panel shows that the
        low latitude structure seen in the other panels is not present at
        Galactic latitude $b \sim -55^\circ$ (though one can see stars from the
	Sagittarius tidal stream scattering into our data at a radial
	velocity of about -160 $\rm km\>s^{-1}$).
}

\setcounter{table}{0}

\begin{deluxetable}{rrcccccrrcr}
\tabletypesize{\scriptsize}
\tablecolumns{11}
\footnotesize
\tablecaption{Blue stars near (l,b) = $(198^\circ, -27^\circ)$ (Table Stub only  -- revised)}
\tablewidth{0pt}
\tablehead{
\colhead{R.A.} &
\colhead{Dec.} &
\colhead{SDSS ID} & \colhead{Fiber ID} &
\colhead{$g_0$} & \colhead{$(g-r)_0$} &
\colhead{$(u-g)_0$} & \colhead{RV} & \colhead{W$_K$} &
\colhead{Select} & \colhead{E(B-V)}\\
\colhead{$^\circ$} & \colhead{$^\circ$} & \colhead{r-re-c-f-id} & \colhead{plate-mjd-fiber} & \colhead{mag} & \colhead{mag} & \colhead{mag} & \colhead{km s$^{-1}$} & \colhead{\AA } & \colhead{Flag} & \colhead{mag} }

\startdata
 70.598290 &  -0.213427 & 0125-7-3-546-0318 & 797-52263-318 & 19.632 &  0.232 &  0.984 &   13.1 &  6.19 & 1  & 0.0606\\
 70.603553 &  -0.237871 & 0125-7-3-546-0219 & 797-52263-313 & 19.919 &  0.280 &  0.731 &   50.0 &  3.36 & 1  & 0.0573 \\
 70.622982 &   0.047201 & 0125-7-4-546-0262 & 797-52263-359 & 19.788 &  0.264 &  0.708 &   98.5 &  5.01 & 1  & 0.0768 \\
 70.625600 &  -0.057840 & 1752-0-3-332-0329 & 797-52263-358 & 19.230 &  0.272 &  1.042 &   66.5 &  4.11 & 1  & 0.0798 \\
 70.652157 &  -0.544781 & 1752-0-2-333-0301 & 797-52263-311 & 19.397 &  0.223 &  0.892 &  192.1 &  3.76 & 1  & 0.0346 \\
 70.788270 &   0.555787 & 0125-7-5-547-0374 & 797-52263-339 & 19.229 &  0.138 &  1.122 &   92.2 &  5.03 & 0  & 0.0788 \\
\enddata
\end{deluxetable}

\begin{deluxetable}{lrrrrrrrrrr}
\tablecolumns{11}
\footnotesize
\tablecaption{Summary of Stream detection pieces -- revised}
\tablewidth{0pt}
\tablehead{
\colhead{Name} & \colhead{$\alpha$} & \colhead{$\delta$} & \colhead{l} & \colhead{b} &
\colhead{$<v_R>$} & \colhead{$\sigma(v_R)$} & \colhead{$g_0$\tablenotemark{1}} &
\colhead{$(g-r)_0$\tablenotemark{1}}& \colhead{$N_{\rm stream}$} &\colhead{$v_{circ}$} }

\startdata
Sp198-27-19.8\tablenotemark{2}& 72 & 0& 198 & -27 & 54 $\pm 5$ & 18 & 19.8 & 0.24 & 138 & 173$\pm 51$ \\
Sp225+28-19.6& 133 & 2& 225 & 28 & 78 $\pm 6$ & 13 & 19.6 & 0.26 & 33 & 225$\pm 29$\\
Sp182+27-19.4& 117 & 38& 182 & 27 & 14 $\pm 5$ & 26 & 19.4 & 0.28 & 115 & ---\\
Sp188+24-19.3& 115 & 32 & 188 & 24 & 19 $\pm 5$ & 24 & 19.3 & 0.28 & 75 & 272$\pm 120$ \\
\hline
Sag. South Strm& 35 & 0&165 & -55 & -160 $\pm 5$\tablenotemark{3} & 22\tablenotemark{3} & 21.5\tablenotemark{3} & 0.22\tablenotemark
{3} & 137 & --- \\

\enddata
\tablenotetext{1}{mag and color of faint blue Monoceros turnoff except where indicated}
\tablenotetext{2}{Plate 797}
\tablenotetext{3}{For the Sagittarius stream, this $<v_R>$ and $\sigma$ refer to Sagittarius stream blue horizontal branch and blue
straggler stars rather than those in the Monoceros structures. The color and mag are those of the Sag. stream.}
\end{deluxetable}


\clearpage

\plotone{f2rev.eps}


\plotone{f6rev.eps}


\plotone{f9rev.eps}


\clearpage
\begin{thebibliography}{}
\bibitem[Allen(2000)]{a00} Allen, C. W. 2000, in Allen's Astrophysical Quantities, ed. A. N. Cox
\bibitem[Alard(2001)]{a01} Alard, C. 2001, preprint astro-ph/0007013
\bibitem[Bahcall \& Soneira(1984)] {bs84} Bahcall, J. N. \& Soneira, R. M. 1984, \apjs, 55, 67
\bibitem[Beers et al.(2002)]{bdrcrfnh2002} Beers, T. C., Drilling, J. S., Rossi, S., Chiba, M., Rhee, J., F\"uhrmeister, B, Norris, J. E., and von Hippel, T. 2002, \aj, 124, 931
\bibitem[Bellazzini, Ferraro, \& Ibata(2002)]{bfi02} Bellazzini, M., Ferraro, F., Ibata, R. 2002, \aj, 124, 915.
\bibitem[Blanton et al.(2002)]{betal02} Blanton, M. R., Lupton, R. H., Maley, F. M., Young, N., Zehavi, I., \& Loveday, J. 2002, AJ, in press
\bibitem[Chen et al.(2001)]{cetal01} Chen, B. et al. 2001, \apj, 553, 184
\bibitem[Dohm-Palmer et al.(2001)]{dpetal01} Dohm-Palmer, R. C., Helmi, A., Morrison, H., Mateo, M., Olszewski, E. W., Harding, P., Freeman, K.C., Norris, J., \& Shechman, S. A. 2001 ApJ 555, L37
\bibitem[Fukugita et al.(1996)]{figdss96} Fukugita, M., Ichikawa,T., Gunn, J. E., Doi, M., Shimasaku, K., Schneider, D. P. 1996, \aj, 111, 1758
\bibitem[Gilmore, Wyse, \& Norris(2002)]{gwn2002} Gilmore, G., Wise, R. F. G., \& Norris, J. E. 2002, \apjl, 574, L39
\bibitem[Gilmore \& Wyse(1985)]{gw85} Gilmore, G., and Wyse, R. F. G. 1985, \aj, 90, 2015
\bibitem[Gunn et al.(1998)]{getal98} Gunn, J. E. et al. 1998, \aj, 116, 3040
\bibitem[Helmi et al.(1999)]{hwzz99} Helmi, A., White, S. D. M., de Zeeuw, P. T., and Zhao, H. 1999, \nat, 402, 53
\bibitem[Hogg et al.(2001)]{hetal01} Hogg, D. W. , Finkbeiner, D. P., Schlegel, D. J., \& Gunn, J. E. 2001, AJ, 122, 2129
\bibitem[Ibata et al.(2001a)]{iils01} Ibata, R., Irwin, M., Lewis, G. F., Stolte, A. 2001a, \apj, 547, L133
\bibitem[Ibata et al.(2001b)]{ilitq01} Ibata, R., Lewis, G. F., Irwin, M., Totten, E., and Quinn, T. 2001b, \apj, 551, 294
\bibitem[Ibata et al.(2003)]{ietal2003} Ibata, R. A., Irwin, M. J., Lewis, G. F., Ferguson, A. M. N., and Tanvir, N. 2003, \mnras, in press (astro-ph/0301067)
\bibitem[Kundu et al.(2002)]{ketal02} Kundu, A., et al. 2002, \apj, 576, L125
\bibitem[Larsen \& Humphreys(1996)]{lh96} Larsen, J. A., \& Humphreys, R. M. 1996, \aj, 468, 99
\bibitem[Lupton et al.(2003)]{l03} Lupton, R. H., et al., \aj, in preparation
\bibitem[Majewski et al.(1999)]{mskrjtlp99} Majewski, S. R., Siegel, M. H., Kunkel, W. E., Reid, I. N., Johnston K. V., Thompson, I. B., Landolt, A. U., and Palma, C. 1999, \aj, 118, 1709
\bibitem[Majewski, Munn, and Hawley(1996)]{mmh96} Majewski, S. R., Munn, J., A., and Hawley, S. L. 1996, \apjl, 459, L73
\bibitem[Majewski(1993)]{m93} Majewski, S.~R.\ 1993, \araa, 31, 575 
\bibitem[Malin \& Carter(1983)]{mc83} Malin, D. F. \& Carter, D. 1983, \apj, 274, 534
\bibitem[Martinez-Delgado et al.(2001)]{magc01} Martinez-Delgado, D., Aparicio, A., Gomez-Flechoso, M. Angeles, \& Carrera, Ricardo 2001, \apj, 549, L199
\bibitem[Mateo, Olszewski \& Morrison(1998)]{mom98} Mateo, M., Olszewski, E. W., \&  Morrison, H. 1998, \apj, 508, L55
\bibitem[Mathieu et al.(1986)]{mlgg86} Mathieu, R. D., Latham, D. W., Griffin, R. F., \& Gunn, J. E. 1986, AJ, 92, 1100
\bibitem[Merrifield \& Kuijken(1998)]{mk98} Merrifield, M. \& Kuijken, K. 1998, \mnras, 297, 1292
\bibitem[Monet, Richstone, \& Schechter(1981)]{mrs81} Monet, D. G., Richstone, D. O, \& Schechter, P. 1981, \apj, 245, 454
\bibitem[Morrison, Flynn, \& Freeman(1990)]{mff90} Morrison, H. L., Flynn, C., \& Freeman, K. C. 1990, \aj, 100, 1191
\bibitem[Newberg, Yanny et al.(2002)]{netal02} Newberg, H., Yanny, B., et al. 2002, ApJ, 569, 245 (Paper II)
\bibitem[Norris(1994)]{n94} Norris, J. E. 1994, \apj, 431, 645
\bibitem[Norris, Bessell, \& Pickles(1985)]{nbp85} Norris, J., Bessell, M. S., \& Pickles, A. J. 1985, \apjs, 58, 463
\bibitem[Odenkirchen et al.(2001)]{oetal01} Odenkirchen, M. et al. 2001, \apjl, 548, L165
\bibitem[Parker, Humphreys, \& Larsen(2001)]{phl2001} Parker, J. E., Humphreys, R. M., and Larsen, J. A. 2001, \baas, 199, \#91.05
\bibitem[Pier et al.(2002)]{pmk03} Pier, J. et al. 2002, \aj, 125, 1559.
\bibitem[Quillen (2002)]{q02} Quillen, A. C. 2002, AJ, 124, 400
\bibitem[Richstone (1980)]{r80} Richstone, D. O. 1980, \apj, 238 , 103.
\bibitem[Rockosi et al.(2002)]{retal02} Rockosi, C. M. et al. 2002, AJ, 124, 349
\bibitem[Schlegel, Finkbeiner, \& Davis(1998)]{sfd98} Schlegel, D.J., Finkbeiner, D.P., \& Davis, M. 1998, ApJ, 500, 525
\bibitem[Smith et al.(2002)]{setal02} Smith, J. A. et al. 2002, AJ, 123, 2121
\bibitem[Stoughton et al.(2001)]{setal01} Stoughton, C., et al. 2001, \aj, 123, 485
\bibitem[Tremaine(1999)]{t99} Tremaine, S. 1999, \mnras, 307, 877
\bibitem[Vivas et al.(2001)]{v01} Vivas, A.~K.~et al.\ 2001, \apjl, 554, L33 
\bibitem[Wilhelm, Beers, \& Gray(1999)]{wbg99} Wilhelm, R., Beers, T., Gray, R. O. 1999, \aj, 117, 2308
\bibitem[Yanny, Newberg et al.(2000)]{ynetal00} Yanny, B., Newberg, H. J., et al. 2000, \apj, 540, 825 (Paper I)
\bibitem[York et al.(2000)]{yetal00} York, D.G.  et al. 2000, AJ, 120, 1579
\end{thebibliography}

\begin{thebibliography}{}
\bibitem[Crane, et al.(2003)]{cmrfsl03} Crane, J. D., Majewski, S. R., Rocha-Pinto, H. J., Frinchaboy, P. M., Skrutskie, M. F., \& Law, D. R. (2003), \apj, 594, L119
\bibitem[Martinez-Delgado et al.(2003)]{mdetal} Martinez-Delgado, D., Gomez-Flechoso, M. A., Aparicio, A., \& Carrera, R., \apj, in press, astro-ph/0308009
\bibitem[Soubiran, Katz, \& Cayrel(1998)]{skc98} Soubiran, C., Katz, D., \& Cayrel, R. (1998), AAS, 133, 221
\end{thebibliography}
\end{document}